\documentclass[showkeys,preprintnumbers,amsmath,amssymb,nofootinbib]{revtex4}
\usepackage{graphicx}
\usepackage{dcolumn}
\usepackage{bm}
\usepackage{amsmath,amssymb,epsfig} 
\usepackage{wrapfig}
\usepackage{comment}

\begin{document}


\title{Advancing Tests of Relativistic Gravity via Laser Ranging to Phobos}

\author{Slava G. Turyshev$^1$, William Farr$^1$, William~M.~Folkner$^1$, Andr\'e~R.~Girerd$^1$, Hamid Hemmati$^1$, Thomas W. Murphy,~Jr.$^2$, James~G.~Williams$^1$, and John J. Degnan$^3$
}

\affiliation{\vskip 3pt
$^1$Jet Propulsion Laboratory, California Institute of Technology,\\
4800 Oak Grove Drive, Pasadena, CA 91109-0899, USA
}%

\affiliation{\vskip 3pt
$^2$Center for Astrophysics and Space Sciences, University of California, 
San Diego,\\
9500 Gilman Drive, La Jolla, CA 92093-0424, USA
}%

\affiliation{\vskip 3pt
$^3$Sigma Space Corporation, 4801 Forbes Blvd., Lanham, MD 20706, USA
}%

\date{\today}

\begin{abstract}
Phobos Laser Ranging (PLR) is a concept  for a space mission designed to advance tests of relativistic gravity in the solar system. PLR's primary objective is to measure the curvature of space around the Sun, represented by the Eddington parameter $\gamma$, with an accuracy of two parts in $10^7$, thereby improving today's best result by two orders of magnitude. Other mission goals include measurements of the time-rate-of-change of the gravitational constant, $G$ and of the gravitational inverse square law at 1.5 AU distances--with up to two orders-of-magnitude improvement for each.  The science parameters will be estimated using laser ranging measurements of the distance between an Earth station and an active laser transponder on Phobos capable of reaching mm-level range resolution. A transponder on Phobos sending 0.25~mJ, 10~ps pulses at 1 kHz, and receiving asynchronous 1~kHz pulses from earth via a 12~cm aperture will permit links that even at maximum range will exceed a photon per second. A total measurement precision of 50~ps demands a few hundred photons to average to 1 mm (3.3~ps) range precision. Existing satellite laser ranging (SLR) facilities--with appropriate augmentation--may be able to participate in PLR. Since Phobos' orbital period is about 8 hours, each observatory is guaranteed visibility of the Phobos instrument every Earth day. Given the current technology readiness level, PLR could be started in 2011 for launch in 2016 for 3 years of science operations. We discuss the PLR's science objectives, instrument, and mission design. We also present the details of science simulations performed to support the mission's primary objectives.
\end{abstract}

\maketitle


\section{\label{sec:intro}Introduction}

Our solar system is a unique laboratory that offers many opportunities to test relativistic gravity.  The true renaissance in tests of general relativity began in the 1970s with major advances in several disciplines--notably microwave spacecraft tracking, astrometric observations with very long baseline interferometry (VLBI), and  lunar laser ranging (LLR). Because of this  technological progress our knowledge of gravity improved significantly  \cite{Turyshev-2008ufn}. 

In particular, analysis of data obtained from radio ranging to the Viking spacecraft determined Eddington's metric parameter\footnote{To describe the accuracy achieved in the solar system experiments, it is useful to refer to the parameterized post-Newtonian (PPN) formalism (see discussion in Sec.~\ref{sec:ppn} and \cite{Will_book93}). Two parameters are of interest here, the PPN parameters $\gamma$ and $\beta$, the values of which in general relativity are $\gamma = \beta=1$. The introduction of $\gamma$ and $\beta$ is useful with regard to measurement accuracies \cite{Turyshev-2008}. In the PPN formalism, the angle of light deflection is proportional to ($\gamma$ + 1)/2, so that astrometric measurements might be used for a precise determination of the parameter $\gamma$. The parameter $\beta$ contributes to the relativistic perihelion precession of a body's orbit.} $\gamma$ at the level of $1.000 \pm 0.002$ \cite{viking_reasen}. Geodetic measurements with VLBI have yielded  value of $\gamma=0.99983\pm0.00045$ \cite{Shapiro_SS_etal_2004}. LLR, a continuing legacy of the Apollo program, provided improved constraint on the combination of parameters $\eta=4\beta-\gamma-3$ \cite{Williams-etal-2004,Williams-etal-2009}. The analysis of LLR data \cite{Turyshev-Williams-2007} constrained this combination as $\eta=(4.0\pm4.3)\times10^{-4}$, leading to an accuracy of $\sim$0.011\% in verification of general relativity via precision measurements of the lunar orbit. Finally, microwave tracking of the Cassini spacecraft on its approach to Saturn improved the measurement accuracy of the parameter $\gamma$ to $\gamma-1=(2.1\pm2.3)\times10^{-5}$, thereby reaching the current best accuracy of $\sim$0.002\% provided by tests of gravity in the solar system \cite{Bertotti-Iess-Tortora-2003} (see discussion in \cite{Turyshev-2008}).

While most of the progress in tests of gravity on interplanetary scales was demonstrated using the microwave frequencies, laser ranging techniques gained significant momentum \cite{Pearlman-etal-ILRS-2002}. Laser ranging from the Earth to passive targets on the lunar surface routinely operates at centimeter level accuracy \cite{Williams-etal-2009}. Millimeter-level accuracy LLR data has been achieved by the Apache-Point Observatory Lunar Laser-ranging Operation (APOLLO) \cite{Murphy-etal-2008,Battat-etal:2008}. Over the years, LLR has benefited from a number of improvements both in observing technology \cite{Pearlman-etal-ILRS-2002} and data modeling. Today LLR is a primary technique to perform high-accuracy tests of relativistic gravity including tests of the Equivalence Principle (EP), of time-variability in the gravitational constant, and of the inverse-square law of gravity.  

\begin{figure}[h!]
 \vspace{-0pt}
  \begin{center}
    \includegraphics[width=0.80\textwidth]{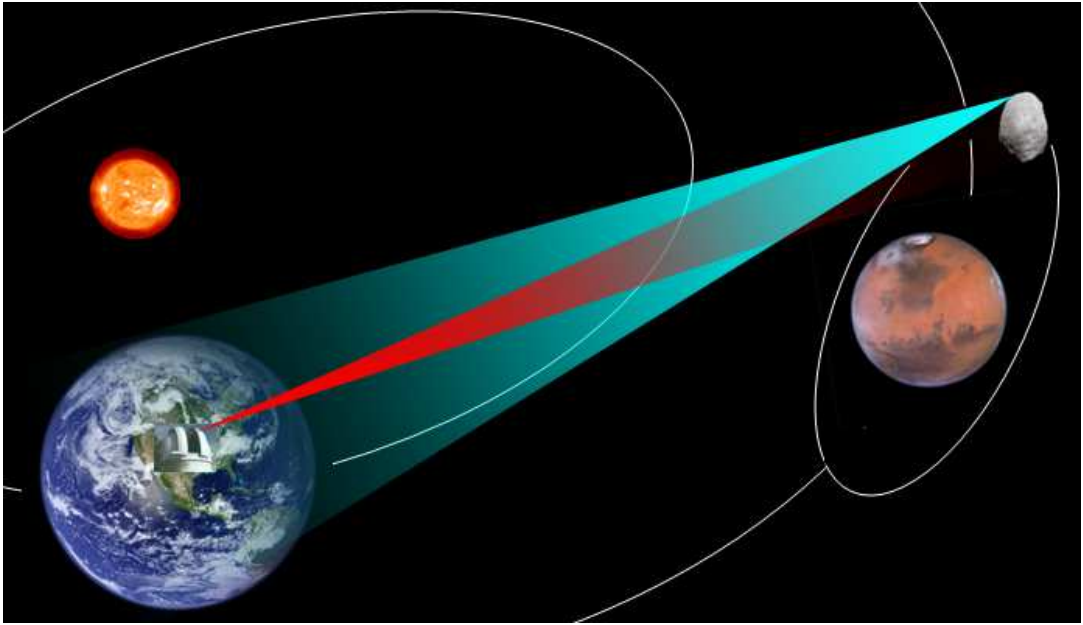}
  \end{center}
  \vspace{-10pt}
  \caption{Concept of a laser transponder link between an observatory on Earth and a laser terminal on Phobos.}
\label{fig:opti-trasnsponder-PLR}
  \vspace{-1pt}
\end{figure}

Building on our experience with LLR data analysis, APOLLO's design, construction and operation, and optical communications technology development, we investigated the science case, and the instrument and mission design of a surface-landed mission to Phobos that would operate a pulsed laser transponder (time-of-flight) capable of achieving mm-level accuracies in ranging between Earth and the orbit of Mars. The resulting Phobos Laser Ranging (PLR) experiment will build on the success of LLR, but will break the passive lunar paradigm (strong signal attenuation due to $1/r^4$ energy transfer) and extend the effectiveness of this technique to interplanetary scales (Figs.~\ref{fig:opti-trasnsponder-PLR}, \ref{fig:plr-concept}). 

At interplanetary distances, active techniques are required to achieve good signal strength (a benefit of $1/r^2$ energy transfer). The development of active laser techniques would extend the accuracies characteristic of passive laser ranging to interplanetary distances \cite{Degnan-2002}.  Technology is available to conduct such measurements, achieving single-photon time resolution measured in tens of picoseconds (ps). One millimeter of range corresponds to 3.3 ps; millimeter range precision can be statistically achieved with a few hundred photons in both uplink/downlink directions. For comparison, several-meter accuracies have been demonstrated with radio tracking at Mars \cite{Konopliv-etal-2006}. 

Interplanetary laser ranging has been recently demonstrated with the MESSENGER (MErcury Surface, Space ENvironment, GEochemistry and Ranging)\footnote{See details on the MESSENGER mission at {\tt http://messenger.jhuapl.edu/}} spacecraft that successfully established a two-way laser link over 24 million kilometers of space between the Earth and the spacecraft, achieving 20 cm range accuracy \cite{Smith-etal-2006,Sun-etal-2006}. In addition, a successful transmission of hundreds of laser pulses over the distance of nearly 81 million km (0.54~AU) from Earth to the Mars Orbiter Laser Altimeter (MOLA)\footnote{See details on the MOLA instrument at {\tt http://mola.gsfc.nasa.gov/}}, an instrument on the Mars Global Surveyor (MGS) spacecraft in orbit about Mars, have demonstrated maturity of laser ranging technologies for interplanetary applications \cite{Abshire-etal-2006}. In fact, the MLA and MOLA experiments demonstrated that decimeter interplanetary ranging is within the state of the art and can be achieved with modest laser powers and telescope apertures. Achieving mm-level precision over interplanetary distances is within reach, thus opening a way to significantly more accurate (several orders of magnitude) tests of gravity on solar system scales \cite{Turyshev-Williams-2007,Degnan-2007a,Degnan-2007b,Birnbaum-Chen-Hemmati-2010}.

The PLR experiment would conduct investigations in several science areas, including gravitational astrophysics and science investigations of Phobos, Mars and the asteroids over a nominal 3 year science mission. Similar analysis of achievable science outcomes for various tests of general relativity was conducted for the radio-science experiments with the future BepiColombo Mercury planetary orbiter  \cite{Milani-etal-2002,Ashby-Bender-Wahr-2007} and also suggested for the interplanetary laser ranging experiments  \cite{Turyshev-Williams-2007,Chandler-etal-2004,Zuber-Smith-2009,Murphy-etal-2009,Hemmati-etal-2009}. In this paper we focus on highly-accurate laser ranging measurements taken on interplanetary scales and their benefits for gravitational physics. 

This paper is organized as follows: Section \ref{sec:2} describes the science objectives of the PLR experiment and significance of anticipated results for gravitational and other fundamental physics.  Section \ref{sec:3.1} discusses the trade space for interplanetary laser ranging. Section \ref{sec:3} presents details of our mission trade studies and results of mission simulations. Section \ref{sec:4} presents a technical overview of the PLR experiment and describes the mission and instrument design, operations and methodology.  We conclude in Section \ref{sec:concl} with a summary and recommendations. 

\section{\label{sec:2}The Science Objectives of the PLR Experiment}
 
Recent progress in observational cosmology has challenged Einstein's general theory of relativity as a model for gravitation in our universe.  From a theoretical standpoint, the challenge is even stronger--if gravity is to be quantized, general relativity will have to be modified.  The continued inability to merge gravity with quantum mechanics, together with the challenges posed by the discovery of dark energy, indicates that the pure tensor gravity of general relativity needs modification or augmentation. It is believed that new physics beyond general relativity and the Standard Model of particles and fields is needed to resolve this issue \cite{Turyshev-2008,Turyshev-etal-2009}.

The kinds of new physics that can solve the problems above typically involve new physical interactions, some of which could manifest themselves as violations of the EP, variation of fundamental constants, modification of the inverse square law of gravity at various distances, and corrections to the current model of spacetime around massive bodies. Each of these manifestations offers an opportunity for experimentation and could lead to a major discovery.

To respond to these challenges, PLR will push high-precision tests of fundamental gravity into a new regime, by using the spacetime in our solar system to explore the physics of the universe. In particular, PLR will improve limits on Eddington's parameterized post-Newtonian (PPN) parameter $\gamma$ \cite{Will-2006}, testing the constancy of Newton's gravitational constant, $G$, searching for a new long range interaction via tests of the Yukawa forces on interplanetary scales, and conducting tests of the EP.  Similarly to its very successful lunar predecessor, PLR could become a major solar-system facility to advance research in fundamental physics; it will also benefit the development of interplanetary optical communication. 

\subsection{\label{sec:2.1.1}What is the nature of spacetime? Does PPN parameter $\gamma$ differ from its general relativistic value? }

Modern gravitational research addresses fundamental questions at the intersection of particle physics and cosmology--including quantum gravity and the very early universe. This work generated ideas on large extra dimensions, large-distance modification of gravity and brane inflation in string theory, all leading to experimentally-testable explanations for the quantum stability of the weak interaction scale. Recent extensions of gravitational models--including brane-world models and efforts to modify models of gravity on large scales--motivate new searches for experimental signatures of small deviations from general relativity on a variety of scales, including solar system distances (see, e.g., \cite{Turyshev-2008}).

Given the immense challenge posed by the unexpected discovery of the accelerated expansion of the universe, it is important to explore every opportunity to explain and probe the underlying physics.  Theoretical efforts in this area offer a rich spectrum of new ideas that can be tested by experiment on solar system scales \cite{Turyshev-etal-2009}.

The Eddington parameter, $\gamma$, whose value in general relativity is unity, is the most fundamental parameter in that $\frac{1}{2}(1-\gamma)$ is a measure, for example, of the fractional strength of the scalar gravity interaction in scalar-tensor theories of gravity \cite{Damour_Nordtvedt_93a,Damour_Nordtvedt_93b,Damour_Piazza_Veneziano_02b}.  Specifically, the quantity $\frac{1}{2}(1-\gamma)$ defines corrections to the spacetime around massive bodies. To date, the most precise value for this parameter, $\gamma-1=(2.1\pm2.3)\times10^{-5}$, was obtained using microwave tracking to the Cassini spacecraft \cite{Bertotti-Iess-Tortora-2003} during a solar conjunction experiment. This accuracy approaches the region where multiple tensor-scalar gravity models, consistent with recent cosmological observations \cite{Spergel:2006hy}, predict a lower bound for the present value of this parameter at the level of $(1-\gamma) \sim 10^{-6}-10^{-7}$.  Therefore, improving the measurement of $\gamma$ would provide crucial information to separate modern scalar-tensor theories of gravity from general relativity, probe possible ways for gravity quantization, and test modern theories of cosmological evolution \cite{Turyshev-etal-2009}.
(In addition, any experiment pushing the present upper bound on another Eddington parameter $\beta$, i.e. $\beta - 1 = (1.0 \pm 1.1) \times 10^{-4}$ from \cite{Williams-etal-2009,Turyshev-Williams-2007}, will also be of interest.)

\begin{figure}[h!]
  \vspace{-0pt}
  \begin{center}
    \includegraphics[width=0.65\textwidth]{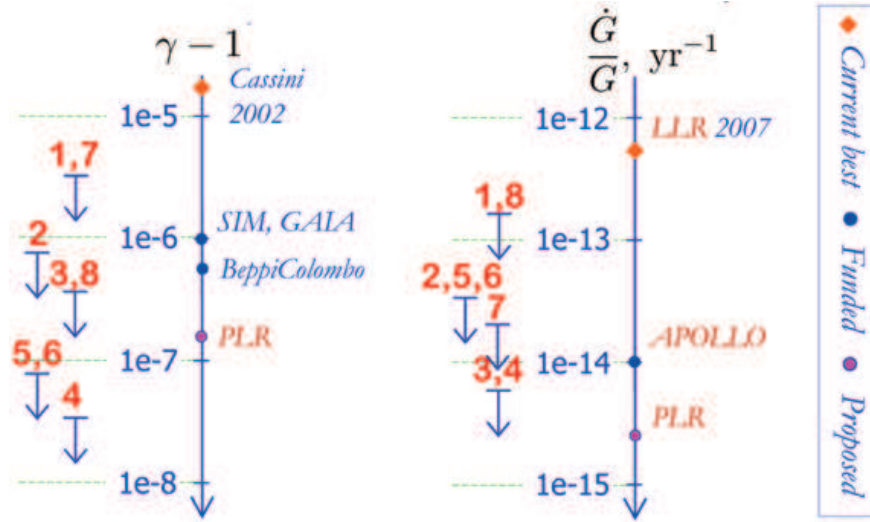}
  \end{center}
  \vspace{-5pt}
  \caption{Scientific discovery space with respect to the PPN parameter $\gamma$ and search for time-variations of the gravitational constant as predicted by several modern theories of gravity (indicated here by numbers \#1--8, see details in the text and in Refs.~\cite{Turyshev-2008ufn,Turyshev-etal-2009}).}
\label{fig:lr-gr-table}
\end{figure}

Figure~\ref{fig:lr-gr-table} shows the anticipated scientific discovery space for PLR in measuring the Eddington parameter $\gamma$, as predicted by a number of modern theories of gravity and gravity augmentation models \cite{Turyshev-etal-2009}, presented here as an example.  In particular, prediction \#1 comes from the string-inspired scalar-tensor theories that were developed in \cite{Damour_Polyakov_94a,Damour_Nordtvedt_93b}, the limit \#2 is from the multi-scalar model \cite{Damour-EspFarese-1996-1,Damour-EspFarese-1996-2}, \#3 is from recent work based on the analysis of scalar-tensor gravities done by \cite{Damour_Piazza_Veneziano_02b}, \#4 is from the theory of large extra dimensions \cite{ADD-1999}, \#5 is from brane-world theories and IR-modified gravity \cite{Dvali-Gabadadze-Porrati-2000,Rubakov:2008nh}, \#6 are from the $f(R)$ gravity models \cite{Sotiriou:2008rp}, \# 7 is from relativistic generalization of Milgrom's Modified Newtonian Dynamics (MOND) theory as developed in \cite{Bekenstein:2004ne}, and, and finally, the limit \# 8 is from the scalar-tensor-vector theory proposed by \cite{Moffat:2005si}. 

\begin{table}[h!]
\begin{center}
\caption{Estimated uncertainties are given in the last three columns for parameters of interest as a function of Phobos lander mission duration, with 1 mm laser ranging once per day with $2^\circ$ SEP cut-off and 67 asteroid mass parameters estimated. Note that data for the gravitational inverse-square law are given for the shortest Earth-Mars ranges of 1.5 AU. (The current best value  of solar $J_2^\odot$ is from \cite{Fienga-etal:2009}, which reported $J_2^\odot=(1.82\pm0.47)\times10^{-7}$ . Note that data from the Venus Express mission were downweighted in our simulations, which resulted in less accurate estimates for $J_2$.)
\label{tab:plr-objectives}}
\vskip 2pt
\begin{tabular}{|c|c|c|c|c|} \hline  
Relativistic Effect &  Current best& 
\multicolumn{3}{c|}{Mission duration / N of conjunctions}\\ \cline{3-5}  
to be studied by PLR &  uncertainty & 1.5 yr / 1 cnj & 3 yr / 2 cnj & 6 yr / 3 cnj\\  \hline \hline 
The Eddington parameter $\gamma$   & $2.3\times10^{-5}$ & $3.1\times10^{-7}$ & $1.4\times10^{-7}$ & $7.9\times10^{-8}$ \\
The Eddington parameter $\beta$   & $1.1\times10^{-4}$ & $4.3\times10^{-4}$ & $1.6\times10^{-4}$ & $9.4\times10^{-5}$ \\
Strong Equivalence Principle, $\eta$   & $4.3\times10^{-4}$ & $1.5\times10^{-3}$ & $2.8\times10^{-4}$ & $8.8\times10^{-5}$ \\
Solar oblateness, $J_2$   & $4.7\times10^{-8}$ & $6.9\times10^{-8}$ & $3.2\times10^{-8}$ & $2.3\times10^{-8}$ \\
Parameter $\dot{G}/G$, yr$^{-1}$  & $7\times10^{-13}$ & $1.7\times10^{-14}$ & $2.8\times10^{-15}$ & $1.0\times10^{-15}$ \\
Gravitational inverse-square law & $2\times10^{-9}$ & $4\times10^{-11}$ & $2\times10^{-11}$& $1\times10^{-11}$ \\
\hline  
\end{tabular} 
\end{center}
\vskip -5pt
\end{table}

Table~\ref{tab:plr-objectives} shows the results of parameter estimation arising from simulated laser ranging to Phobos over 1-6 years of operation based on daily 1 mm accuracy range points (see details in Sec.~\ref{sec:3.4.2}). Estimated parameters include orbital elements and masses of the planets and Phobos, up to 67 individual asteroids, and densities for 3 classes incorporating another 230 asteroids. Based on 3 yr simulations, PLR will be able to measure the PPN parameters $\gamma$ and $\beta$ with accuracy levels of $2\times10^{-7}$ and $2\times10^{-4}$, respectively, to measure the solar oblateness coefficient $J_2$ with an accuracy of $3\times10^{-8}$, to measure the fractional time-rate-of-change of the gravitational constant to $3\times10^{-15}$~yr$^{-1}$, and to test the gravitational inverse square law to an accuracy of $2\times10^{-11}$ at $\sim$1.5 AU distances and the strong EP parameter to $3\times10^{-4}$. Thus, PLR could directly verify important predictions of modern theories with an unprecedented accuracy, enabling strong advances in the tests of general relativity. More simulation details will be discussed in Sec.~\ref{sec:3.4.2}.

\subsection{\label{sec:2.1.2}Do the fundamental constants of Nature vary with space and time?}

The possibility that fundamental physical parameters may vary with space and time has been revisited with the advent of models unifying the forces of nature based on the symmetry properties of extra dimensions, such as Kaluza-Klein-inspired theories, Brans-Dicke theory, and supersymmetry models.  Modern gravitational models that attempt to ``complete'' general relativity at very short distances (embedding it in a more powerful theory capable of addressing phenomena on that scale) or ``extend'' it on very large distances ($\sim10^{28}$ cm) typically include cosmologically evolving scalar fields that lead to variability of the fundamental constants.  

A variation of the cosmological scale factor with epoch could lead to temporal or spatial variation of the gravitational constant, $G$. A possible variation of $G$ could be related to the expansion of the universe depending on the cosmological model considered.  Variability in $G$ can be tested in space with a much greater precision than on Earth \cite{Williams-etal-2004}.  For example, a decreasing gravitational constant, $G$, coupled with angular momentum conservation is expected to increase a planet's semimajor axis, $a$, as $\dot{a}/a=-\dot{G}/G$.  The corresponding change in orbital longitude grows quadratically with time, providing for strong sensitivity to the effect of $\dot{G}/G$.

Lunar and planetary ranging measurements currently lead the search for very small spatial or temporal gradients in the value of $G$ \cite{Williams-etal-2004,Muller-Biskupek-2007}.  Recent analysis of LLR data \cite{Williams-etal-2004,Turyshev-Williams-2007} strongly limits such variations and constrains a local ($\sim$1 AU) scale expansion of the solar system as $ \dot{a}/a=-\dot{G}/G=-(6\pm7)\times10^{-13}$ yr$^{-1}$, including those stemming from cosmological effects. Interestingly, the achieved accuracy in $\dot{G}/G$ implies that, if this rate is representative of our cosmic history, then $G$ has changed by $\leq$1\% over the 13.7~Gyr age of the universe.  The ever-expanding LLR data set and its increasing accuracy will lead to further improvements in the search for spatial or temporal variability of $G$.  PLR represents the next advance in testing for variations in $G$, see Figure~\ref{fig:lr-gr-table}. 

With 1-mm ranging accuracy, PLR data will permit a measurement of $\dot{G}/G$  to $\sim3\times10^{-15}$ yr$^{-1}$ in about three years, see Table~\ref{tab:plr-objectives}.  This anticipated accuracy is a result of our simulations that account for the range perturbations due to 67 large asteroids. 

\subsection{\label{sec:2.1.3}Do extra dimensions or other new physics alter the inverse square law (ISL)?}

Many modern theories of gravity, including string, supersymmetry, and brane-world theories, suggest that new physical interactions will appear at various ranges thereby modifying Newton's gravitational ISL.  While most attention has focused on the behavior of gravity at short distances (for a review of experiments, see \cite{Adelberger-Heckel-Nelson-2003}), it is possible that non-compact extra dimensions could produce small deviations from the ISL on interplanetary scales \cite{Dvali-Gruzinov-Zaldarriaga-2003,Turyshev-etal-2007,Turyshev-etal-2009}. 
An experimental confirmation of new fundamental for\-ces would provide an important insight into the physics beyond the Standard Model.

\begin{figure}[h!]
  \vspace{-0pt}
  \begin{center}
    \includegraphics[width=0.60\textwidth]{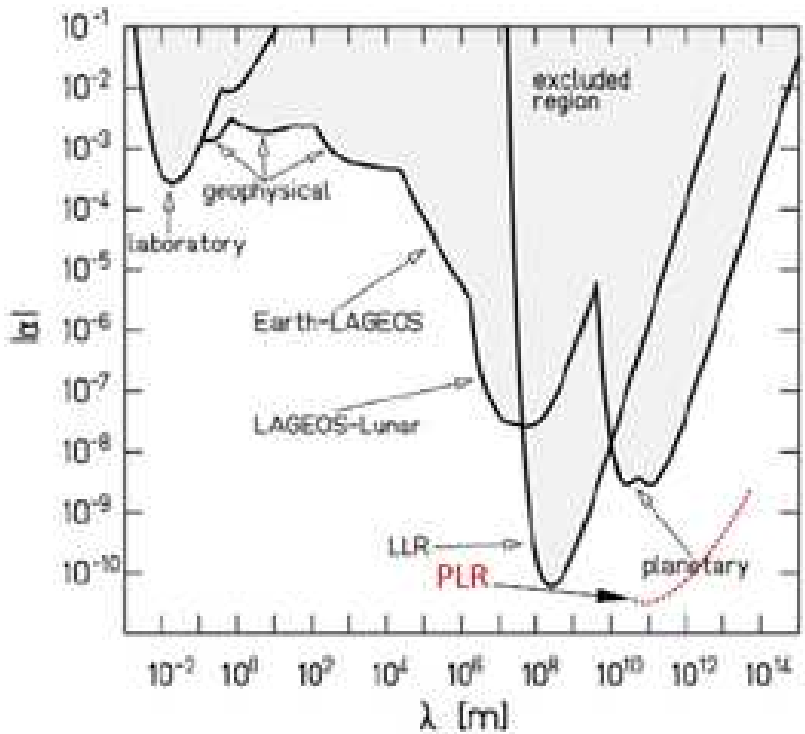}
  \end{center}
  \vspace{-5pt}
  \caption{Limits on the ISL violations \cite{Adelberger-Heckel-Nelson-2003}; PLR will improve ISL tests in outer solar system.  Parameters $\alpha$ and $\lambda$ are respectively the strength and range of a new physical interaction.}
\label{fig:isl-tests-plr}
\end{figure}

In its simplest versions, a new interaction (or a fifth force) would arise from the exchange of a light boson coupled to matter with a strength comparable to gravity. Planck-scale physics could give origin to such an interaction in a variety of ways, thus yielding a Yukawa-type modification in the interaction energy between point-like masses. In general, the interaction energy, $V(r)$, between two point masses $m_1$ and $m_2$ can be expressed in terms
of the gravitational interaction as
{}
\begin{equation} 
V(r) = -  {G_{\infty}m_{1}m_{2} \over r}\big(1 +
\alpha\,e^{-r/\lambda}\big), \label{eq:2.1} 
\end{equation}
\noindent where $r = \vert {\bf r}_2 - {\bf r}_1 \vert$ is the distance between the masses, $G_{\infty}$ is the gravitational coupling for $r \rightarrow \infty$ and $\alpha$ and $\lambda$ are respectively the strength and range of the new physical interaction. Naturally, $G_{\infty}$ has to be identified with Newton's gravitational constant and the gravitational coupling becomes dependent on $r$ \cite{Turyshev-2008ufn}. Indeed, the force associated with Eq.~(\ref{eq:2.1}) is given by: ${\bf F}(r)={\vec \nabla} V(r)=  
 - G(r)m_{1}m_{2}\,\hat {\bf r}/r^2$,  where effective gravitational constant $G(r)$ has the form 
$G(r) = G_{\infty}\big[1 + \alpha\,(1 + r/\lambda)e^{-r/\lambda}\big]$.

By far the most stringent constraints on violation of the ISL at \emph{large distances} to date come from very precise measurements of the lunar orbit about the Earth. Analysis of the LLR data tests the gravitational ISL to $3\times10^{-11}$ of the gravitational field strength on scales of the 385,000 km Earth-moon distance \cite{Muller-etal-2006}.  New LLR facilities \cite{Murphy-etal-2008} will reach sensitivity of $10^{-12}$ at Earth-moon distances, while mm-level laser ranging to Phobos could improve the corresponding ISL solar system tests by two orders of magnitude at interplanetary scales (see Fig.~\ref{fig:isl-tests-plr}).

\subsection{\label{sec:2.1.4}Is the Equivalence Principle exact?}

Einstein's Equivalence Principle (EP) lies at the foundation of the general theory of relativity; testing this fundamental assumption with the highest possible sensitivity is clearly important, particularly since it is expected that the EP will not hold in quantum theories of gravity. The Equivalence Principle can be split into two parts: the weak equivalence principle (WEP) tests the sensitivity to composition and the strong equivalence principle (SEP) checks the dependence on mass.   In its \emph{strong form} (the SEP) the EP covers the gravitational properties resulting from gravitational energy itself, thus involving an assumption about the non-linear property of gravitation \cite{Williams-etal-2009}. In the SEP case, the relevant test body differences are the fractional contributions to their masses by gravitational self-energy. Because of the extreme weakness of gravity, a test of the SEP requires bodies with astronomical sizes.  Although general relativity assumes that the SEP is exact, many modern theories of gravity typically violate the SEP by including new fields of matter, notably scalar fields \cite{Damour_Nordtvedt_93a,Damour_Nordtvedt_93b,Damour_Polyakov_94a,Damour_Polyakov_94b}.

Currently, the Earth-moon-Sun system provides the best solar system arena for testing the SEP. LLR experiments involve bouncing laser beams off retro-reflector arrays placed on the moon during the period 1969 to 1973 \cite{Williams-etal-2009}. A violation of the EP leads to a radial perturbation of the lunar orbit that, for the SEP, is $\delta r \sim (4\beta-3-\gamma)~13 \cos D$ meters \cite{Nordtvedt_1968a,Nordtvedt_1968b,Damour_Vokrouhlicky_1996,Nordtvedt-2003}, with $\beta$ and $\gamma$ being two PPN parameters. LLR investigates the SEP by looking for a displacement of the lunar orbit along the direction to the Sun. There are laboratory investigations of the WEP (at University of Washington) which are about as accurate as LLR  \cite{Baessler_etal_1999,Adelberger_2001}. Recent solutions using LLR data yield a test of the EP of $\Delta (M_G/M_I)_{\tt EP} =(-0.8\pm1.3)\times10^{-13}$ (corrected for solar radiation pressure as suggested in \cite{Vokrouhlicky-1997}), where $\Delta (M_G/M_I)$ signifies the difference between gravitational-to-inertial mass ratios for the Earth and the moon. After adjusting for laboratory WEP tests  \cite{Nordtvedt-2003,Baessler_etal_1999,Adelberger_2001}, this result corresponds to a test of the SEP of $\Delta (M_G/M_I)_{\tt SEP} =(-1.8\pm1.9)\times10^{-13}$ with the SEP violation parameter $\eta=4\beta-\gamma-3$ found to be $\eta=(4.0\pm4.3)\times 10^{-4}$  \cite{Williams-etal-2004,Williams-etal-2009,Turyshev-Williams-2007}. Further improvements may be achieved via mm-precision ranges to the moon \cite{Turyshev-Williams-2007,Murphy-etal-2008} and new laser ranging instruments deployed on the moon or other planetary surfaces.

A lander on Phobos can push these tests a bit further.  In particular, the Sun-Earth-Phobos-Jupiter system tests the SEP in a qualitatively different way from LLR \cite{Anderson-etal-1996}.  The SEP polarization effect is much larger for the Earth and Mars orbits than for the lunar orbit, but the optimum time span is longer than the nominal 3 yr mission.  With 1 mm precision ranging, the EP-violating polarization toward Jupiter allows the SEP violating parameter $\eta$ to be measured to $3\times10^{-4}$ for observations spanning up to three years.  Additional improvement will be possible with an extended duration PLR mission. 

In summary, the proposed PLR mission will provide tests of relativistic gravity in the solar system to an unprecedented precision.  It will test the weak-gravity-and-small-speed regime of the cosmologically motivated theories that explain the small acceleration rate of the Universe (a.k.a. dark energy) via modification of gravity at cosmological scales.  PLR would search for a cosmologically-evolved scalar field that is predicted by modern theories of quantum gravity and cosmology, and also by superstring and brane-world models \cite{Turyshev-2008,Turyshev-etal-2009}.  With an accuracy of two parts in ten million anticipated from PLR measuring the Eddington parameter $\gamma$, this mission could discover a violation or extension of general relativity, and/or reveal the presence of any additional long range interaction.

\section{\label{sec:3.1}Trade Space for Interplanetary Laser Ranging}

Highly-accurate time-series of the travel times of laser pulses between an observatory on the Earth and an optical transponder on a solar system object could lead to major advances in several science areas, including tests of general relativity, solar system dynamics, and studies of the target body. Any solar-system object with a transparent atmosphere and a solid surface would be a suitable platform to deploy a laser ranging instrument \cite{Chandler-etal-2004}. 

The main objective of our study was to identify a solar system location that would enable a highly accurate measurement of the gravitational time delay (given by Eq.~(\ref{eq:light-time})), which is our primary science signal. Such a location should provide the conditions necessary for a much improved determination of the PPN parameter $\gamma$ together with a number of other relativistic gravity parameters, as shown in Table~\ref{tab:plr-objectives}.  

The planet Mars is one of those solar-system bodies that are well suited for gravitational physics experiments. Proximity to the Earth and the maximization of signal strength is certainly a major factor in choosing an Earth-Mars link. Furthermore, the recent increase of space traffic on the Earth-Mars route makes this location quite feasible for this purpose. Therefore, it is quite natural to consider Mars as a location of the remote laser terminal as part of an interplanetary laser ranging experiment that has the potential to improve the values of several relativistic parameters \cite{Anderson-etal-1996,Turyshev-Williams-2007}.  
However, there are factors that make this location less desirable compared to others.     

As part of the effort to optimize the mission concept, a trade space of laser ranging locations was explored.  The search was limited to celestial bodies within the solar system, since single element solar orbiting ``free-flyer'' spacecraft were considered outside the desired mission cost category due to the perceived need for their drag-free characteristics with attendant high expense.  The trade space was bounded by the basic concept of laser ranging between Earth and another planetary location that could offer opportunities for solar conjunctions with laser beams passing near the sun giving good $\gamma$ sensitivity.  Mission concepts outside this fundamental construct were not included in the analysis.

Guided by our initial expectations, we started our studies with Mars being the target body for the experiment. However, in recent studies we investigated a trade space of mission alternatives, as opposed to an effort focusing exclusively on a narrow refinement of the original mission concept.  This approach bore two major results.  First, one of the alternatives explored proved superior to the original concept.  Since the Phobos-based mission proved to be equivalent in science and lower in both cost and risk compared to the original Mars-based mission, it became the preferred solution.  The second result of the trade space exploration was a strong understanding of the suitability of major alternative locations within the solar system for ranging-based fundamental gravity science.  This provided a solid, defensible basis for the conclusion that Phobos is the optimal planetary location for laser ranging to and from Earth, given the targeted scientific goals and mission class.

\begin{figure}[h!]
  \vspace{-5pt}
  \begin{center}
    \includegraphics[width=0.96\textwidth]{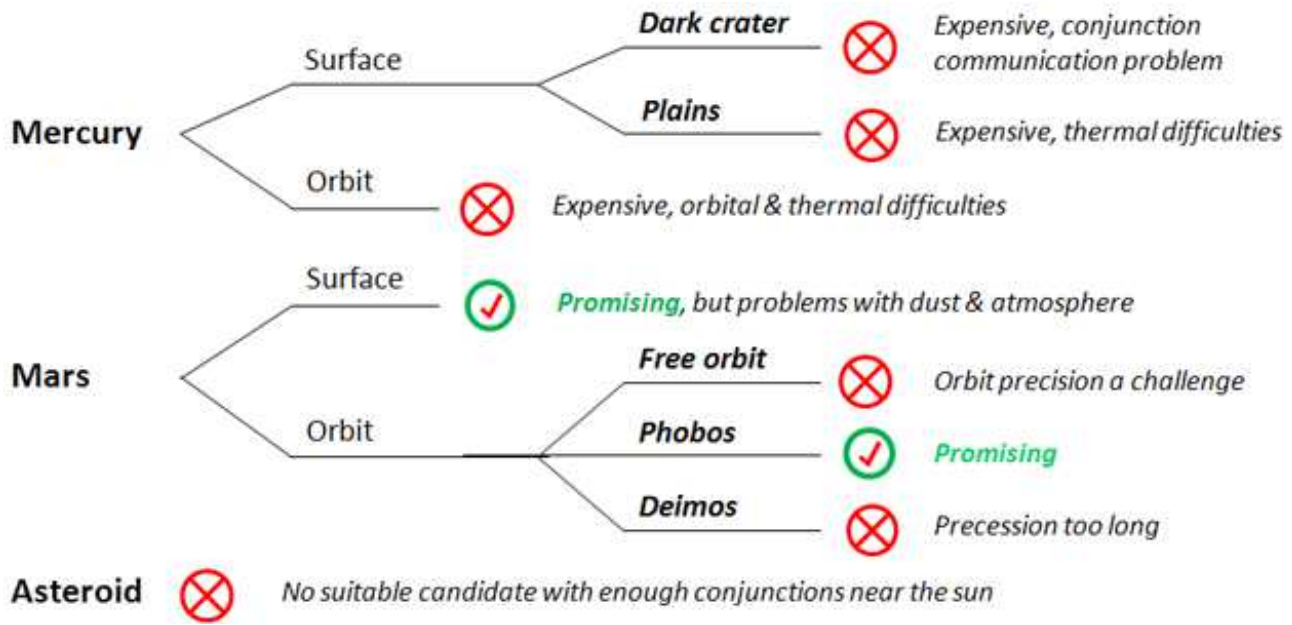}
  \end{center}
  \vspace{-10pt}
  \caption{Interplanetary laser ranging trade space explored in the study.}
\label{fig:trade-space}
\end{figure}

The high-level results of the trade space exploration are shown in Figure~\ref{fig:trade-space}.  The trade space included Mercury (both orbiters and landers), Mars (orbiters and landers) and asteroids. Although Mercury offered a competitive science return, the practical challenges of surviving in its environment for any meaningful duration made it infeasible, or at least too expensive to remain viable; also, time and cost of orbit transfer were more significant compared to other location alternatives.  Due to significant influence of non-gravitational forces affecting the spacecraft motion (solar radiation pressure, thruster firing, etc.) orbital candidates suffered orbit precision difficulties.  Suitable asteroid candidates could not be found that exhibited enough science-rich near-sun conjunctions over reasonable mission lifetimes to stay competitive.  The two winners warranting more investigation were the original Mars surface location and Phobos. Phobos was preferred to Deimos as a site for laser ranging: because of its faster orbit, shorter precession period (Phobos' orbital precession period is  2.26 yr, while for Deimos this period is 54 yr), and smaller area-to-mass ratio. 

Below we present results of a further down-select that resulted in choosing Phobos as the preffered location for the laser ranging mission. 

\subsection{\label{sec:3.3}Tradeoffs between Phobos and Mars}

Both Mars and Phobos laser ranging missions were studied in detail; overall Phobos is a better choice. Mars is dynamically well suited to studying gravitational physics, but ranging through its dusty atmosphere presents a challenge and solar panels must be large enough to compensate for dust accumulation. The 8 hours (hr) Phobos orbit and rotation periods offer three tracking opportunities per day. Our simulations show (see Sec.~\ref{sec:3}) that ranging to Phobos has good sensitivity to improve the values of several gravitational physics parameters. Phobos edged out Mars as the preferred solution due to a superior science/cost ratio at lower mission risk.  
The extra parameters needed to model Phobos' orbit and orientation accurately are no more daunting than dealing with Mars orientation.
Detailed comparisons between the two concepts are provided as follows:

\begin{itemize}
\item Phobos landing is not complicated by atmospheric entry and descent, so Entry, Descent, and Landing (EDL) is simpler, but landing consequently needs more $\Delta V$. It is approximately 2 km/sec more using multiple burns to match orbits. Mars landings are familiar but risky. 
\item With current technology, Mars landing sites must be low altitude since atmospheric breaking is used. That shifts latitudes north causing greater seasonal variations of sunlight and temperature, plus longer nights and lower Earth elevation angles in winter. Phobos landing sites can be equatorial with less seasonal variation and better seasonal elevation angle variation of Earth and Sun.
\item Mars has windblown dust that scatters light making the sky a bright noise source while dimming the signal, and dust settles on optics and solar panels. There is no dusty atmosphere on Phobos, but electrostatically levitated dust is possible, though less trouble. The Phobos sky would be dark with no absorption. 
\item A spacecraft on Phobos would likely last longer than on Mars, since it would not experience dust on the solar panels, cold winter nights, and reduced winter power from lower elevation of the winter Sun. 
\item Ranging to a lander on Mars is sensitive to precession, nutation, UT1\footnote{UT1 is the principal form of Universal Time (UT), see details on various timescales at {\tt http://en.wikipedia.org/wiki/Universal\_Time}} and polar motion, which orient Mars \cite{Yoder-Standish-1997,Folkner-etal-1997a,Folkner-etal-1997b}. Precession is sensitive to moment of inertia, and nutation is sensitive to interior structure. UT1 and polar motion are affected by the changing polar caps. UT1 is also sensitive to tides. 
\item Phobos orbit and physical librations add dynamical complexity to the range model, but those complexities trade off with Mars UT1, polar motion, nutations, and geocenter motion. Mars UT/PM has been dealt with during previous lander mission, while Phobos physical libration has barely been detected and lacks detailed measurement. 
\item The Phobos orbit has a strong tidal acceleration in longitude, but this should be tractable. Phobos would be affected by Mars gravity, but the gravity field has been determined by orbiting spacecraft. It is benficial to orbit Phobos on arrival in order to map its gravity field. Phobos has a much larger area/mass ratio than Mars which would make it much more vulnerable to impact impulses. 
\item Since Phobos makes three orbits per day, a single station on the Earth can always track during one, and sometimes two, orbits. By contrast, the similarity of Mars and Earth rotation periods interferes with daily tracking from one Earth site requiring at least two tracking sites at different Earth longitudes for daily coverage.
\item 4 hr nights on Phobos, 1/3 of Mars, may require less stored power. But there are three times as many temperature cycles. 
\item For radio relay through Mars orbiting spacecraft, a Phobos lander would need to be on the hemisphere facing Mars. During part of the Mars year Phobos experiences occultations by Mars as seen from Earth and the Sun experiences eclipses by Mars. These occultations and eclipses can be up to 54 minutes long so that as little as 2.9 hours per orbit, about 9 hours/Earth day, may be available for ranging and charging batteries, and the missing times may be for Earth and Sun high in the sky depending on landing longitude. There would be good coverage when the Earth and Sun are furthest north or south such as local winter; the eclipses and occultations occur for $|$Mars referenced declinations$|<21^\circ$ (Mars obliquity is $25^\circ$). There would be increased thermal cycling during the two eclipse seasons each Mars year. Eclipses and occultations affect only the Mars facing hemisphere of Phobos. 
\item There is no Phobos atmospheric range correction required and thus a meteorological package is unnecessary.
\item It is easier to support a line of sight close to the sun from the surface of Phobos (due to lack of atmospheric scintillation) than from Mars--a clear advantage in accessing the region from which sensitivity to the primary PPN $\gamma$ signal is greatest.
\item Communication via radio relay is likely less efficient on Phobos than on Mars due to relay directionality, so direct communication with Earth may be preferred. Both sides of Phobos (i.e., Mars-facing and far-side) could be used for this purpose. 
\item Phobos was preferred to Deimos as a site for laser ranging because of its faster orbit, shorter precession period, and smaller area-to-mass ratio.
\end{itemize}

The considerations above resulted in our choice of Phobos as opposed to Mars as our primary target body for the interplanetary laser ranging experiment.  

\section{\label{sec:3} Phobos Laser Ranging Mission Simulations}

In this section we discuss the details of the simulations performed to support the PLR's primary science objectives presented in Sec.~\ref{sec:2}.  

\subsection{Design of the sensitivity studies}
\label{sec:design_sensitivity}

In order to assess the potential science return from laser ranging to a suitably equipped lander, numerical simulations were performed. The simulations were based on the Solar System Data Processing Software developed at JPL and used for the development of planetary ephemerides for NASA missions. The planetary ephemeris (the most recent of which is DE421 \cite{Folkner-etal-2008}) is based on a large variety of measurements including radio range measurements to Mars landers (Viking, Pathfinder) and orbiters (MGS, Odyssey, MRO). The software includes the capability to model and accurately estimate the measured round-trip distance and its dependence on a wide variety of dynamical parameters.

The orbits of the planets and asteroids are integrated based on the Einstein-Infeld-Hoffmann equations of motion derived for the PPN metric for general relativity \cite{Einstein-Infeld-Hoffmann-1938,Moyer-2003} (see Eq.~\ref{eq:4-26-mod}). The orbits are integrated based on nominal values for the orbital elements of the planets, Sun, moon, and asteroids, their mass parameters, and PPN parameters. The sensitivities to these parameters are evaluated by re-integrating the equations of motion many times with perturbations of the initial conditions by small amounts from the nominal for each parameter taken one at a time. This procedure generates partial derivatives from small finite differences of orbits with respect to initial conditions, masses, PPN parameters, etc. Estimates of the dynamical parameters are made using a wide variety of observational data in a global least-squares fit (e.g. \cite{Folkner-etal-2008}).

\subsection{The PPN equations of motion}
\label{sec:ppn}

Generalizing on a phenomenological parameterization of the gravitational metric tensor field, which Eddington originally developed for a special case, a method called the PPN formalism has been developed (see \cite{Will-1973,Will_book93} for details). This method represents the gravity tensor's potentials for slowly moving bodies and weak inter-body gravity and is valid for a broad class of metric theories, including general relativity as a unique case. The several parameters in the PPN metric expansion vary from theory to theory, and they are individually associated with various symmetries and invariance properties of the underlying theory  (see \cite{Will_book93} for details).

If (for the sake of simplicity) one assumes that Lorentz invariance, local position invariance and total momentum conservation hold, the point-mass Newtonian and relativistic perturbative accelerations in the solar system's barycentric frame has the following form:
{}
\begin{eqnarray}
\ddot{\bf r}_i&=&\sum_{j\not=i}\frac{\mu_j({\bf r}_j-{\bf r}_i)}{r_{ij}^3}\bigg\{
1-\frac{2(\beta+\gamma)}{c^2}\sum_{l\not=i}\frac{\mu_l}{r_{il}}-
\frac{2\beta-1}{c^2}\sum_{k\not=j}\frac{\mu_k}{r_{jk}}+
\nonumber\\
&&+~
\gamma\big(\frac{{\dot r}_i}{c}\big)^2+(1+\gamma)\big(\frac{{\dot r}_j}{c}\big)^2-\frac{2(1+\gamma)}{c^2} (\dot{\bf r}_i \hskip-1pt\cdot\hskip-1pt \dot{\bf r}_j)-\nonumber\\
&&-~\frac{3}{2c^2}\bigg[\frac{({\bf r}_i-{\bf r}_j)\hskip-1pt\cdot\hskip-1pt{\dot{\bf r}}_j}{r_{ij}}\bigg]^2+\frac{1}{2c^2}({\bf r}_j-{\bf r}_i)\hskip-1pt\cdot\hskip-1pt{\ddot{\bf r}}_j\bigg\}+
\nonumber\\
&&+~
\frac{1}{c^2}\sum_{j\not=i}\frac{\mu_j}{r_{ij}^3}
\Big\{\Big[{\bf r}_i-{\bf r}_j\Big]\hskip-1pt\cdot\hskip-1pt\Big[ (2+2\gamma){\dot {\bf r}}_i-(1+2\gamma){\dot {\bf r}}_j\Big]\Big\}({\dot{\bf r}}_i-{\dot{\bf r}}_j)+
\nonumber\\
&&+~\frac{3+4\gamma}{2c^2}\sum_{j\not=i}\frac{\mu_j{\ddot {\bf r}}_j}{r_{ij}}+{\cal O}(c^{-4}),
\label{eq:4-26-mod}
\end{eqnarray}
\noindent where the indices $j$ and $k$ refer to the $N$ bodies and where $k$ includes body $i$, whose motion is being investigated. $\mu_j$ is the gravitational constant for body $j$ given as $\mu_j=Gm_j$, where $G$ is the universal Newtonian gravitational constant and $m_j$ is the isolated rest mass of a body $j$. In addition, the vector ${\bf r}_i$ is the solar system barycentric radius-vector of this body, the vector ${\bf r}_{ij}={\bf r}_j-{\bf r}_i$ is the vector directed from body $i$ to body  $j$, and $r_{ij}=|{\bf r}_j-{\bf r}_i|$. The dimensionless parameters $\gamma$ and $\beta$ are the two Eddington PPN parameters. General relativity, when analyzed in standard PPN gauge, gives $\gamma=\beta=1$; other theories yield different values of these parameters \cite{Will_book93}.  

To account for a possible violation of the EP and temporal changes in the gravitational constant one can add another perturbative acceleration term $\delta\ddot{\bf r}_i$:
{}
\begin{eqnarray}
\delta\ddot{\bf r}_i&=&\sum_{j\not=i}\frac{\mu_j({\bf r}_j-{\bf r}_i)}{r_{ij}^3}\bigg\{
 \Big(\left[\frac{m_G}{m_I}\right]_i-1\Big)
+\frac{{\dot G}}{G}\cdot (t-t_0)\bigg\},
\label{eq:4-26-mod-PPN}
\end{eqnarray}
\noindent where parameter $([{m_G}/{m_I}]_i-1)$ signifies a possible inequality between the gravitational and inertial masses that is needed to facilitate investigation of a possible violation of the EP (see Sec.~\ref{sec:2.1.4}). In addition, Eq.~(\ref{eq:4-26-mod-PPN}) also includes parameter ${\dot G}/{G}$, which is needed to investigate a possible temporal variation in the gravitational constant (see Sec.~\ref{sec:2.1.2}). Note that in general relativity $\delta\ddot{\bf r}_i\equiv0$.

The accuracy of the relativity tests with interplanetary laser ranging will depend on our knowledge of the solar gravity field. A source of  uncertainty in these tests is the dimensionless solar quadrupole moment $ J_{2\odot}$. The value of $ J_{2\odot}$ from solar oscillation data has been reported as $(1.7\pm0.4)\times10^{-7}$ by \cite{Duvall-etal-1984} and as about 10\% less than this value by \cite{Brown-etal-1989} derived from the oblateness of the photosphere. These results are consistent with the value expected for nearly uniform rotation of the Sun. However, solar oscillation data suggest that most of the Sun rotates slightly slower than the surface except possibly for a more rapidly rotating core \cite{Duvall-Harvey-1984,Godier-Rozelot-2001} suggesting the value of $(2.0\pm0.4)\times 10^{-7}$. Recently Fienga et al. reported determination of $J_2^\odot$ at the level of $(1.82\pm0.47)\times10^{-7}$ using interplanetary ranges including Venus Express \cite{Fienga-etal:2009}. Thus the solar oblateness is not an issue for the laser ranging tests of gravity with a lander on Phobos/Mars (although $J_2^\odot$ may be important for the case of ranging to a Mercury lander). The accuracy achievable from interplanetary laser ranging may improve upon these results. For this reason, and because of the expected correlation of $J_2$ with $\beta$ and other parameters, it is necessary to include $J_2$ in the solution parameter list. The acceleration of a test body in the field of the sun must then be modified to read

\begin{equation}
\ddot{\bf a}_i = \ddot{\bf r}_i +\delta\ddot{\bf r}_i-
\frac{GM_\odot}{2}J_2\nabla\Big[\frac{R_\odot^2}{r_{i\odot}^3}(3\sin^2\phi-1)\Big]
\label{eq:total}
\end{equation}
where $R_\odot$ is the radius of the Sun and $\phi$ is the latitude of
the planet with respect to the solar equator. The terms $\ddot{\bf r}_i$ and $\delta\ddot{\bf r}_i$ are given by Eqs.~(\ref{eq:4-26-mod}) and (\ref{eq:4-26-mod-PPN}) correspondingly. 

To determine the orbits of the planets and spacecraft one must also describe propagation of electro-magnetic signals between the two points in space.  The corresponding one-way light-time equation, with accuracy sufficient for the purposes of this paper, can be presented as below 
\begin{equation}
\label{eq:light-time}
t_2-t_1= \frac{r_{12}}{c}+(1+\gamma)\sum_i \frac{\mu_i}{c^3}
\ln\left[\frac{r_1^i+r_2^i+r_{12}}{r_1^i+r_2^i-r_{12}}\right]+{\cal O}(c^{-5}),
\end{equation}
\noindent  where $t_1$ refers to the signal transmission time at body 1, and $t_2$ refers to the reception time at body 2.  $r^i_{1}=|{\bf r}^i_{1}(t_1)|$ and $r^i_{2}=|{\bf r}^i_{2}(t_2)|$ are the positions of the transmitter and receiver with respect to the $i$-th body (including the sun) at transmission and reception times correspondingly, and $r_{12}=|{\bf r}_2(t_2)-{\bf r}_1(t_1)|$ is their spatial separation.  
The logarithm term in Eq.~(\ref{eq:light-time}) provides our principal sensitivity to parameter $\gamma$. We emphasize that the measurement of this PPN parameter constitutes the primary objective of the proposed mission (see Sec.~\ref{sec:2.1.1}). 

The PPN expansion serves as a useful framework to test relativistic gravitation in the context of the gravitational experiments. The main properties of the PPN metric tensor are well established and are widely used in modern astronomical practice \cite{Moyer-1981-1,Moyer-1981-2,Standish_etal_92,Will_book93}. The equations of motion Eq.~(\ref{eq:total}) are used to produce numerical codes for the purposes of constructing the solar system's ephemerides, spacecraft orbit determination \cite{Moyer-2003,Standish_etal_92}, and analysis of the gravitational experiments in the solar system \cite{Will_book93,Turyshev-2008}. In our analysis we also used this set of equations of motion.

\subsection{Estimated parameters}
\label{sec:3.4par}

In this section, we discuss a series of studies designed to illustrate the results that might be obtained with an interplanetary laser ranging (ILR) transponder on Mercury, Mars and Phobos. We used the same data set as in DE421 \cite{Folkner-etal-2008}, except for the LLR data and planetary transits. The set of estimated parameters included the set of seven parameters, namely the value of the astronomical unit (AU), the PPN parameters $\gamma$ and $\beta$, solar $J_{2\odot}$, the time-rate-of-change in the gravitational constant ${\dot G}/{G}$, $GM_{\rm Earth}/GM_{\rm moon}$, and the EP parameter $([{m_G}/{m_I}]_\odot-1)\simeq\eta U_\odot$, with $U_\odot=-3.52\times 10^{-6}$ being the gravitational binding energy of the Sun \cite{Ulrich-1982,Anderson-etal-1996}. For inverse-square law simulations, a Yukawa amplitude was estimated for a fixed length.

\begin{table}[h!]
\begin{center}
\vskip -0pt
\caption{Typical set of solar-system parameters estimated in the study (in addition to those listed in Table~\ref{tab:plr-objectives}). Note that in addition to 8 planetary bodies, the orbital elements for Pluto were also estimated. \label{tab:est-param}}
\vskip 2pt
\begin{tabular}{|r|c|} \hline  
Estimated parameters  
& \multicolumn{1}{c|}{Number}\\ \hline \hline 
Planetary orbital elements & 54 \\
Astronomical unit (AU) & 1 \\
$GM_{\rm Earth}/GM_{\rm moon}$ & 1\\
Asteroid GMs & 67 \\
Asteroid class densities & 3 \\
Spacecraft biases & 8\\
Solar corona corrections & 4  \\
Mercury shape & 8 \\
Venus radius & 1 \\
Mars Lander locations & 9\\
\hline
\multicolumn{1}{r}{Total number:}
&\multicolumn{1}{c}{156}\\
\end{tabular} 
\end{center}
\vskip -15pt
\end{table}

Our studies combined the current solar-system data set with the simulated ILR data in a simultaneous covariance analysis. In addition to the six Table~\ref{tab:plr-objectives} parameters of interest to gravitational physics, Table~\ref{tab:est-param} shows a summary of the parameters in our model of the solar system. The total set of the estimated solar-system parameters included 162 parameters. The 8 spacecraft biases used in the study and reported in Table~\ref{tab:est-param} are for the two Viking landers 1 and 2 (VL1, VL2), Mars Pathfinder (MPF), Mars Global Surveyor (MGS), Odyssey (ODY), Mars Reconnaissance Orbiter (MRO), Mars Express (MEX), and Venus Express (VEX). In addition, 4 parameters used to estimate the solar corona corrections came from the analysis of the radio-metric tracking data received from MGS, ODY, MRO, MEX spacecraft.  
Note that data from Venus Express mission were downweighted in our simulations, which resulted in less accurate estimates for $J_2$ presented in Tables~\ref{tab:plr-objectives}, \ref{tab:C1}-\ref{tab:C4}.

Initial simulations were done for a laser transponder on a Mars lander. The range from an Earth station to a Mars lander requires knowledge of the locations of the station and lander, the orientation of Earth and Mars as a function of time, the orbits of Earth and Mars including perturbations by the other planets and asteroids, and possible variations in the underlying theory of gravity. 

Later simulations were done for a landed transponder on Phobos. The simulations 
also estimated additional body-specific parameters. In particular, the set of Phobos-specific parameters included
Phobos' dynamic parameters (from integrated Phobos/Deimos ephemeris):
Phobos' initial state (position and velocity),
Mars' $GM$ (separate from planetary list) and tidal lag angle $\psi$ (related to $k_2/Q$),
Phobos' $GM, J_2,$ and $C_{22}$, 
Deimos' initial state (position and velocity), and 
Deimos' $GM$ \cite{Jacobson-2009}. We also included Phobos' geometric parameters, namely bias for lander, lander coordinates,
libration in longitude amplitude (geometric), and 
libration in latitude amplitude and phase.

For the simulations using a spacecraft lander on Phobos, the sensitivity to gravitational physics parameters comes from the 1-yr orbit of the earth-moon system and the 2-yr-orbit of the Mars system. We did not consider the sensitivity to gravitational physics parameters of the 8 hr Phobos orbit around Mars. Tracking of a lander on Phobos is used as a way to follow the motion of more massive Mars through the solar system.

\subsection{\label{sec:3.4}Details of simulation and results}

\subsubsection{\label{sec:3.4.1}Lander on Mars}

For the Mars lander laser ranging mission, simulated measurements between a Mars lander and an Earth station were formed based on a nominal position of the lander and tracking station. Since range measurements within a single hour-long observation pass are expected to be correlated (due to pre-pass instrument calibrations), only a single ``normal'' range point was used for each day. These measurements were then included with the current planetary ephemeris data set to estimate parameter uncertainties, with variations depending on assumptions of data accuracy and duration. The position of the lander was always estimated from the simulated laser range data along with the orbit and Mars rotation parameters. The position of the Earth station was assumed to be known from GPS or another surveying technique.

The orientation of the Earth as a function of time is monitored to the mm level by combinations of VLBI, GPS, and laser ranging data. The orientation of Mars as a function of time is modeled in series form with certain parameters estimated to fit Mars radio range and Doppler data as part of the planetary ephemeris fit \cite{Folkner-etal-1997b,Konopliv-etal-2006}.

Asteroids have a significant effect on the orbit of Mars. For the JPL planetary ephemeris, 67 asteroid mass parameters with the largest effect on Mars were individually modeled while 230 other asteroids with smaller effects were modeled by grouping into three taxonomy classes. In order to investigate the sensitivity of the simulated measurements to asteroid modeling, solutions were formed based on allowing various number of asteroid mass parameters to be completely unconstrained and requiring estimation from the laser ranging data. 

\begin{table}[h!]
\begin{center}
\vskip -0pt
\caption{Estimated uncertainties for parameters of interest as a function of number of estimated asteroid mass parameters, for a 36 month Mars lander mission with 1 mm laser ranging once per day with $2^\circ$ SEP cut-off. \label{tab:C1}}
\vskip 5pt
\begin{tabular}{|c|c|c|c|} \hline  
Relativistic Effect  & 
\multicolumn{3}{c|}{Number of asteroid $GM$s}\\ \cline{2-4}  
to be studied by PLR &  11 & 36 & 67\\  \hline \hline 
The Eddington parameter $\gamma$  & $7.8\times10^{-8}$ & $1.1\times10^{-7}$ & $1.4\times10^{-7}$ \\
The Eddington parameter $\beta$   & $6.9\times10^{-5}$ & $9.7\times10^{-5}$ & $1.7\times10^{-4}$ \\
Strong Equivalence Principle, $\eta$   & $4.3\times10^{-5}$  & 
$8.8\times10^{-5}$ & $3.1\times10^{-4}$ \\
Solar oblateness, $J_2$   & $1.6\times10^{-8}$ & $2.5\times10^{-8}$ & $3.2\times10^{-8}$ \\
Parameter $\dot{G}/G$, yr$^{-1}$  & $2.6\times10^{-15}$ & $2.6\times10^{-15}$ & $2.8\times10^{-15}$ \\
\hline  
\end{tabular} 
\end{center}
\vskip -15pt
\begin{center}
\vskip -0pt
\caption{Estimated uncertainties for parameters of interest as a function of Mars lander mission duration, with 1 mm laser ranging once per day with $2^\circ$ SEP cut-off and 67 asteroid mass parameters estimated. \label{tab:C2}}
\vskip 1pt
\begin{tabular}{|c|c|c|c|} \hline  
Relativistic Effect  & 
\multicolumn{3}{c|}{Mission duration, months}\\ \cline{2-4}  
to be studied by PLR &  18 & 36 & 72\\  \hline \hline 
The Eddington parameter $\gamma$  & $3.1\times10^{-7}$ & $1.4\times10^{-7}$ & $7.8\times10^{-8}$ \\
The Eddington parameter $\beta$   & $4.3\times10^{-4}$ & $1.7\times10^{-4}$ & $8.6\times10^{-5}$ \\
Strong Equivalence Principle, $\eta$   & $8.8\times10^{-4}$  & 
$3.1\times10^{-4}$ & $8.5\times10^{-5}$ \\
Solar oblateness, $J_2$   & $6.9\times10^{-8}$ & $3.2\times10^{-8}$ & $2.1\times10^{-8}$ \\
Parameter $\dot{G}/G$, yr$^{-1}$  & $1.7\times10^{-14}$ & $2.8\times10^{-15}$ & $1.0\times10^{-15}$ \\
\hline  
\end{tabular} 
\end{center}
\vskip -20pt
\end{table}

Table~\ref{tab:C1} shows the estimated uncertainties in the PPN $\gamma$ and $\beta$ parameters, the second zonal harmonic of the solar gravity field, and the rate of change of the gravitational constant $G$. The estimated uncertainties are clearly dependent on the number of asteroid mass parameters that must be estimated. The parameters of most interest, PPN $\gamma$ and $\dot{G}/G$, are less dependent on the asteroid modeling. Since significant improvement in asteroid mass parameters may be achieved through such observations and the main parameters of interest for the simulations are not too strongly dependent on the asteroid modeling, the estimation of 67 main asteroid mass parameters with the laser ranging data is considered to be sufficient for this study. Asteroid mass parameters can also be estimated using observations of the deflection of smaller asteroids that happen to pass near larger asteroids. The number of asteroids observed in this way is increasing \cite{Baer-Chesley-2008}. Also, asteroid masses obtained with other techniques are a benefit.

The potential science return, in terms of uncertainties in estimated parameters, is also dependent on mission duration. Table~\ref{tab:C2} shows the estimated uncertainties in key parameters as a function of mission duration. This is particularly true for the PPN $\gamma$ parameter, which is most strongly measured at solar conjunctions, occurring at the Earth-Mars synodic period of 780 days. For a nominal Type-II trajectory,\footnote{For details, see {\tt http://www2.jpl.nasa.gov/basics/bsf4-1.php}} landing on Phobos is possible a few months before conjunction. A second solar conjunction requires science mission duration of about 3 years, while a mission lifetime of 6 years allows observations through three conjunctions. 

\subsubsection{\label{sec:3.4.2}Lander on Phobos}

Simulations were performed for a lander on Phobos. A Phobos lander avoids issues of dust on the Mars lander. For the Phobos lander the orbit and orientation of Phobos must be estimated along with the planetary dynamic parameters. A numerically integrated orbit of Phobos has been fit to a variety of data by \cite{Jacobson-2009}. Sensitivities with respect to the initial Phobos orbital elements and Mars tidal parameter were obtained and corrections estimated in the overall least-squares fit. To account for possible unmodeled orbital effects, a correction to the orbital elements was estimated each year. Phobos rotation was modeled based on a series by \cite{Borderies-Yoder-1990,Jacobson-2009}. Corrections to the libration amplitudes in longitude and in latitude were estimated to account for necessary updates to the model series. Because the Phobos orbit and rotation have periods of 8 hours, the relevant parameters are mostly uncoupled with the Mars orbital and gravitational parameters of main interest. The simulations performed for Phobos (see Table~\ref{tab:plr-objectives}) showed essentially the same results for gravitational physics parameters as shown in Table~\ref{tab:C2} for the Mars lander case.

Currently the orbit of Mars is modeled to cm-level accuracy to fit radio range measurements with meter-level accuracy, while the orbit of Phobos is modeled to meter-level accuracy to fit spacecraft Doppler data for Phobos fly-bys from Viking, Mars Express, etc. at the 100 m level. A model of Phobos' rotation, intended to be precise to the 5 cm level \cite{Borderies-Yoder-1990}, has been calculated analytically. For processing mm-accurate laser ranges, the modeling for Phobos will be updated building on lunar experience. The rotation state will be explicitly integrated, and forces affecting the orbit of Phobos upgraded to include more accurate tide and solar pressure models. Our PLR simulations include estimates of physical parameters needed to achieve the required modeling accuracy, such as 
the tidal dissipation, Phobos $J_2$ and $C_{22}$, and libration amplitude and phase, and perturbations by asteroids on the Martian orbit. (Note that estimated uncertainties for parameters of interest as a function of number of estimated asteroid mass parameters, for a 36 month Phobos lander mission with 1 mm laser ranging once per day with $2^\circ$ SEP cut-off produced results nearly identical to those for Mars, shown in Table~\ref{tab:C1}.)

Table~\ref{tab:plr-objectives} shows the results of parameter estimation arising from simulated laser ranging to Phobos over 1.5-6 years of operation based on daily 1 mm accuracy range points. Estimated parameters include orbital elements and masses of the planets and Phobos, up to 67 individual asteroids, and densities for 3 classes incorporating another 230 asteroids. Based on 3 yr simulations, PLR will be able to measure the PPN parameters $\gamma$ and $\beta$ with accuracy levels of $2\times10^{-7}$ and $2\times10^{-4}$, respectively, to measure the solar oblateness coefficient $J_2$ with an accuracy of $3\times10^{-8}$, to measure the fractional time-rate-of-change of the gravitational constant to $3\times10^{-15}$~yr$^{-1}$, and to test the gravitational inverse square law to an accuracy of $2\times10^{-11}$ at $\sim$1.5 AU distances and the SEP parameter to $3\times10^{-4}$. Thus, PLR could directly verify important predictions of modern theories with an unprecedented accuracy, enabling strong advances in the tests of general relativity.

\begin{figure}[h!]
  \vspace{-0pt}
  \begin{center}
    \includegraphics[width=0.65\textwidth]{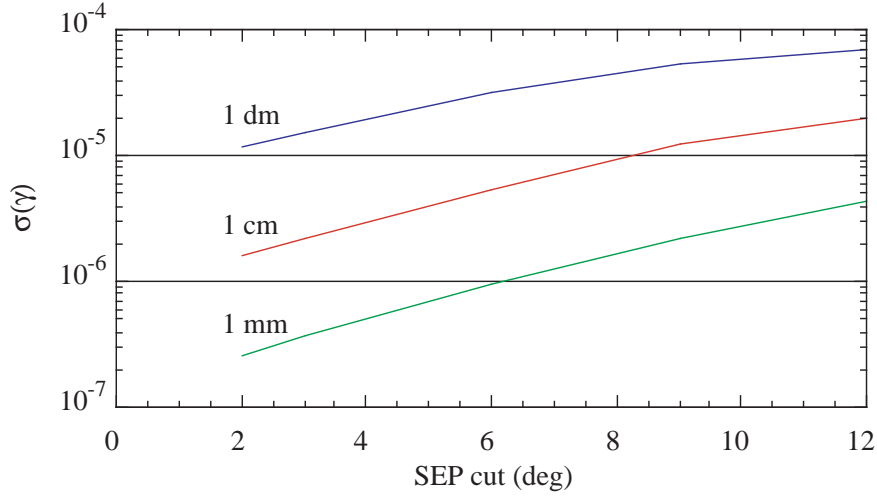}
  \end{center}
  \vspace{-5pt}
  \caption{Estimated uncertainty in PPN $\gamma$ as a function of data accuracy and data cut-off with angular separation from the Sun as viewed from Earth.}
\label{fig:plr-gamma}
\end{figure}

An initial set of simulations were performed for a single conjunction as functions of the data accuracy and angular separation of Mars from the Sun as viewed from Earth. The primary parameter of interest was the PPN $\gamma$, which due to the time delay sensitivity, is not as dependent on orbital uncertainties as other parameters of interest. The results for the estimated uncertainty in PPN $\gamma$ are shown in Figure~\ref{fig:plr-gamma}, which are similar to previous results by \cite{Chandler-etal-2004}. Since the main science goal is to estimate PPN $\gamma$ with an uncertainty of two parts in $10^7$, these results drive the instrument accuracy to 1 mm and require observations within $2^\circ$ of the Sun (as viewed from Earth).

\subsubsection{\label{sec:3.4.3}Lander on Mercury}

Mercury was also investigated as a possible location for the laser ranging lander. With its short orbital period, Mercury provides many more solar conjunction opportunities than for a comparable mission duration at Mars. Mercury is also more sensitive to solar oblateness, and potentially less sensitive to unknown asteroid mass parameters since it is farther from the asteroid belt. Simulations were performed along the same lines as the Mars and Phobos simulations. Table~\ref{tab:C3} shows estimated parameter uncertainties of an 18-month long Mercury surface duration for three different sets of estimated asteroid mass parameters. The sensitivity to asteroid modeling is higher than might be expected, probably because Earth is perturbed at a measurable level by the larger asteroids, and the laser ranging is between Mercury and Earth. Table~\ref{tab:C4} gives results as a function of mission duration. As expected the estimated uncertainties are smaller than in the comparable Mars/Phobos cases, although for the primary parameters of interest, PPN $\gamma$ is improved only by a factor of 3 for the nominal 36-month duration case, and $\dot{G}/G$ is not significantly improved (compare with Table~\ref{tab:C2}).  

\begin{table}[h!]
\begin{center}
\vskip -0pt
\caption{Estimated uncertainties for parameters of interest as a function of number of estimated asteroid mass parameters, for an 18 month Mercury lander mission with 1 mm laser ranging once per day with $2^\circ$ SEP cut-off. \label{tab:C3}}
\vskip 2pt
\begin{tabular}{|c|c|c|c|} \hline  
Relativistic Effect  & 
\multicolumn{3}{c|}{Number of asteroid $GM$s}\\ \cline{2-4}  
to be studied by PLR &  11 & 36 & 67\\  \hline \hline 
The Eddington parameter $\gamma$  & $6.7\times10^{-8}$ & $8.0\times10^{-8}$ & $9.8\times10^{-8}$ \\
The Eddington parameter $\beta$   & $4.2\times10^{-6}$ & $7.3\times10^{-6}$ & $1.5\times10^{-5}$ \\
Strong Equivalence Principle, $\eta$   & $2.6\times10^{-5}$  & 
$5.1\times10^{-5}$ & $1.5\times10^{-4}$ \\
Solar oblateness, $J_2$   & $4.4\times10^{-10}$ & $7.9\times10^{-10}$ & $1.7\times10^{-9}$ \\
Parameter $\dot{G}/G$, yr$^{-1}$  & $1.2\times10^{-14}$ & $1.4\times10^{-14}$ & $1.7\times10^{-14}$ \\
\hline  
\end{tabular} 
\end{center}
\vskip -15pt
\begin{center}
\vskip -0pt
\caption{Estimated uncertainties for parameters of interest as a function of Mercury lander mission duration, with 1 mm laser ranging once per day with $2^\circ$  SEP cut-off and 67 asteroid mass parameters estimated. \label{tab:C4}}
\vskip 2pt
\begin{tabular}{|c|c|c|c|} \hline  
Relativistic Effect  & 
\multicolumn{3}{c|}{Mission duration, months}\\ \cline{2-4}  
to be studied by PLR 
&  18 & 36 & 72\\  \hline \hline 
The Eddington parameter $\gamma$  & $9.8\times10^{-8}$ & $4.6\times10^{-8}$ & $2.7\times10^{-8}$ \\
The Eddington parameter $\beta$   & $1.5\times10^{-5}$ & $6.6\times10^{-6}$ & $7.5\times10^{-7}$ \\
Strong Equivalence Principle, $\eta$   & $1.5\times10^{-4}$  & 
$3.4\times10^{-5}$ &  $6.5\times10^{-6}$ \\
Solar oblateness, $J_2$   & $1.7\times10^{-9}$ & $6.9\times10^{-10}$ & $7.4\times10^{-11}$ \\
Parameter $\dot{G}/G$, yr$^{-1}$  & $1.7\times10^{-14}$ & $2.8\times10^{-15}$ & $9.1\times10^{-16}$ \\
\hline  
\end{tabular} 
\end{center}
\vskip -15pt
\end{table}

\subsection{\label{sec:2.2}Towards PLR Dynamical Model and Secondary Science Objectives}

\begin{figure}[h!]
  \vspace{-0pt}
  \begin{center}
    \includegraphics[width=0.67\textwidth]{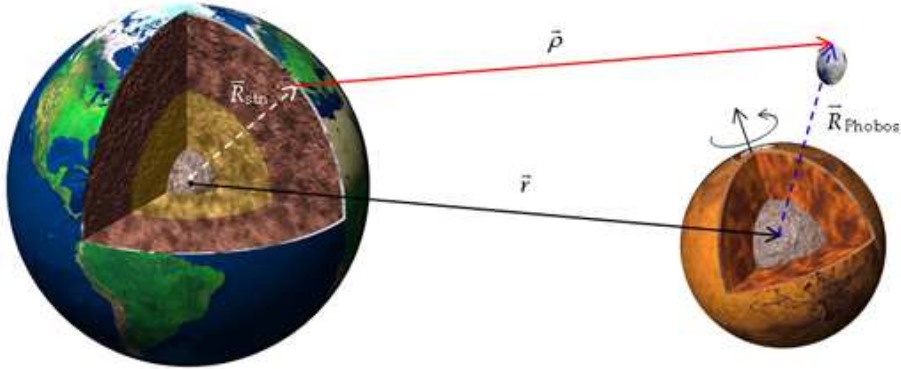}
  \end{center}
  \vspace{-10pt}
  \caption{Precision laser ranging to Phobos could measure the distance between an observatory on the Earth and a terminal on the surface of Phobos to accuracy of 1 mm in less then 5 min of integration time.}
\label{fig:plr-concept}
  \vspace{-0pt}
\end{figure}

The development of the PLR data analysis techniques will build on the successful ongoing lunar and planetary data analysis. Analysis of PLR data requires modeling the orbital and rotational motions of the lander on Phobos and tracking stations on Earth to mm accuracy (see Fig.~\ref{fig:plr-concept}). Models for LLR are currently at cm accuracy for processing range measurements to retroreflectors on the moon. This modeling experience gives confidence that modeling of Phobos laser ranging can reach the mm level. The LLR data and models are used to determine the orbit and rotation of the moon, including sensitivity to lunar interior parameters, and establish the state-of-the art for several fundamental physics parameters such as the upper bound on the violation of the Strong Equivalence Principle (SEP) parameter $\eta$, the PPN parameter $\beta$, the value of the temporal variation in the gravitational constant, and the test of the inverse-square law of gravity \cite{Turyshev-Williams-2007}.

A high quality dynamical model of Phobos rotating and orbiting Mars while Mars orbits the Sun is necessary for analyzing millimeter quality range data to the PLR spacecraft. Some mission concerns and model items deserve special mention and there are also opportunities for science in addition to the gravitational physics objectives. These are described below. 

\subsubsection{\label{sec:2.2.2}Phobos Orbital Science}

Prior to landing on Phobos, the spacecraft would orbit near Phobos for several weeks imaging the surface, measuring topography and sampling the gravity field (see discussion in Sec.~\ref{sec:4.1}). While the primary purpose of this inspection is to provide information to enable a successful landing, there would be a science benefit to mapping surface properties. A good determination of the gravity field would aid the development of a physical libration model and recovery of moments of inertia. A comparison of topography and gravity field would test whether the density is homogeneous. Phobos information has been collected by \cite{Kuzmin-etal-2003}. 

\subsubsection{\label{sec:2.2.3}Phobos Physical Librations}

The orientation of Phobos changes with time. A high accuracy physical libration model will be necessary for successful range data analysis. The largest departure from uniform rotation has an anomalistic period with an amplitude of $\sim1^\circ$ predicted from the shape. A $1^\circ$ amplitude would displace a lander at a surface point opposite Mars by ~230 m. An amplitude of $0.78^\circ \pm0.4^\circ$ has been reported \cite{Duxbury-Callahan-1989}. Measurement of the amplitude will provide the moment of inertia combination $(B-A)/C$, where the moments of inertia satisfy $A<B<C$. In combination with Phobos' gravity coefficient $C_{22}$ from the mission's orbital inspection, it will be possible to determine the moment of inertia $C$. Another physical libration effect, the equator of Phobos tilts a few seconds of arc with respect to the orbit plane. This tilt is sensitive to $(C-A)/B$; the combination $J_2+2C_{22}$ gives $(C-A)$ while $(J_2-2C_{22})$ gives $(C-B)$. A theory of Phobos physical librations is given by \cite{Borderies-Yoder-1990,Chapront-Touze-1990}. Numerically integrated physical librations would need to be developed for millimeter data analysis. 

\subsubsection{\label{sec:2.2.4}Phobos Orbit}

At a first approximation Phobos can be described as orbiting Mars with a precessing slightly elliptical $(e=0.015)$, slightly inclined $(i=1.1^\circ)$ orbit with semimajor axis $a=9375$~km \cite{Jacobson-2009}. The orbit period is 7.654 hr and the mean motion $n$ is 3.136 revolutions/day, faster than Mars' rotation. The major cause of the precession of perifocus and node directions is the oblate figure of Mars. The precession periods are 2.26 yr prograde for perifocus and 2.26 yr retrograde for node. Precession periods shorter than the mission duration benefit separation of parameters during solutions. Small additional perturbations arise from Mars' gravity field (other than the oblate figure), the Sun and Deimos. An orbit theory has been derived by \cite{Chapront-Touze-1988}. 

Perturbations from solar radiation pressure, Mars static and variable gravity field, and tides deserve special discussion below. We also note that while Phobos' orbit plane precesses along Mars' equator plane, that equator plane is precessing and nutating with time. The precession of Mars' equator is of interest because, in combination with Mars' $J_2$, it provides $C/MR^2$ 
\cite{Yoder-Standish-1997,Folkner-etal-1997a,Yoder-etal-2003,Konopliv-etal-2006,Konopliv-etal-2009}. 

\subsubsection{\label{sec:2.2.5}Solar Radiation Pressure}

Solar radiation pressure acting on Phobos displaces the orbit. The acceleration from solar pressure and thermal re-radiation depends on the projected Phobos area divided by its mass.\footnote{The values for the area-to-mass ratio for the moon, Phobos and Mars were evaluated to be $(A/M)_{\tt moon} \simeq 1.29\times 10^{-10}~{\rm m}^2{\rm kg}^{-1}$ for the moon, $(A/M)_{\tt Phobos} \simeq 3.3\times 10^{-8}~{\rm m}^2{\rm kg}^{-1}$ for Phobos, and $(A/M)_{\tt Mars} \simeq 5.65\times 10^{-11}~{\rm m}^2{\rm kg}^{-1}$ for Mars.} 
This effect has been studied for the Earth's moon because it causes a perturbation similar to that from a violation of the equivalence principle \cite{Nordtvedt-1995,Vokrouhlicky-1997}. Unlike the moon, the projected area of Phobos varies about $\pm20$\%. A spherical Phobos would be perturbed by 6.4 mm (vs 3.65 mm for the moon) in orbit radius and about twice that in orbital longitude. The period is the 0.319 d synodic period with respect to the Sun. The real non-spherical Phobos would add harmonics, and the elliptical Mars orbit would cause a $\pm19$\% modulation of the radial amplitude which would vary from 5 to 8 mm with the 687 d Mars orbit period. While the effect of Phobos' shape could get complicated in detail, the relatively small size of the perturbation suggests that a smooth shape model would suffice for the acceleration model. 

Mars' obliquity is $25.2^\circ$. When the Sun is near either of Mars' equinoxes, there will be eclipses of Phobos by Mars and also of Mars by Phobos. This can complicate the solar radiation pressure perturbations of Phobos' orbit by allowing a small accumulation of perturbation in orbit longitude, perhaps as much as a few centimeters in several months. When the Sun is near either solstice, there are no eclipses. Solar radiation pressure and thermal re-radiation could be modeled with four parameters: an acceleration parameter proportional to the area/mass ratio, a thermal lag parameter, and two shape parameters. It should be possible to model solar pressure and thermal re-radiation effects reasonably well and to solve for the model parameters.

\subsubsection{\label{sec:2.2.6}Mars Gravity}

The gravity field of Mars has been mapped with increasing accuracy by successive orbiting spacecraft. The spinning planet is oblate and the unnormalized $J_2$ is $1.96\times10^{-3}$. The large $J_2$ causes the Phobos orbit node and perifocus directions to precess by $159 ^\circ$/yr, retrograde for the node and prograde for the perifocus. While Phobos' eccentricity and inclination are small, these precession rates would be determined to very high accuracy by Phobos laser ranging. Since the sensitivity to higher-degree harmonics falls off as powers of 1/2.76 (the ratio of the radius of Mars $R$ to Phobos' semimajor axis $a$), spacecraft-derived static fields for Mars should be adequate in higher degrees, but it may be necessary to solve for some lower degree gravity coefficients.

Of particular interest are time variations in Mars' gravity field. Seasonal changes of the polar caps, with exchange of CO$_2$ between the poles and variable atmospheric pressure, cause observed variations in the gravity field \cite{Yoder-etal-2003,Konopliv-etal-2006,Smith-etal-2009}. Seasonal changes in $J_3$ are clearly seen from the gravity variations while possible variation in $J_2$ is closer to the noise. Variation in $J_2$ is proportional to variation in the polar moment of inertia, which is independently observed through the rotation of Mars determined from radio tracking of landed spacecraft \cite{Yoder-Standish-1997,Folkner-etal-1997b} and orbiting spacecraft \cite{Konopliv-etal-2006,Konopliv-etal-2009}. $J_4$ and $J_5$ variations are also reported by \cite{Smith-etal-1999}. Phobos has a low inclination orbit so it is most affected by the even zonal coefficients. A $1\times10^{-9}$ variation of the unnormalized $J_2$ with 687 d period would cause a term with 8 m amplitude in Phobos' orbital longitude. The peak-to-peak seasonal variation in $J_2$ predicted by \cite{Smith-etal-1999} is $7\times10^{-9}$. Using this simulation and the rotation information for $J_2$ variations, orbit perturbations with periods of 687 d and 343 d should be prominent (one to several meters). Higher harmonics should be present with smaller amplitudes. Seasonal $J_2$ changes, combined with other even seasonal zonal variations, should be strongly detected by PLR. 

Accurately tracking a lander on Phobos offers great promise for determining effects from the seasonal waxing and waning of the Martian polar caps. At some level, these seasonal changes will not exactly repeat from year to year and that will lead to unpredictable dynamical perturbations of Phobos' orbit. Seasonal changes and small nonperiodic variations in $J_2$ will have to be part of the solution. We do not expect that a small long time scale wander of the orbital longitude of Phobos, which would affect the range observable at short period, would limit the science goals of long period in the center of mass motion of Mars and its satellites around the Sun since that center of mass can be accurately determined over short time spans. Simulations, with the Phobos orbit independently determined every year, successfully recovered the gravitational physics parameters.  

\subsubsection{\label{sec:2.2.7}Tides on Mars}

Solid-body tides are raised on Mars by both Phobos and the Sun. Tides from Phobos are about 1 mm in height while Sun-raised tides are $\sim1$ cm. Gravitational variations from tides on Mars can perturb the orbit of Phobos. Sun-raised zonal tides have long periods, 687 d and 343 d, from orbit eccentricity and obliquity respectively. The corresponding Phobos orbit perturbations are about 1 m in longitude. There are also smaller tidal signatures at 1/3 and other fractions of the 687 d Martian year. The periods are the same as the polar cap $J_2$ variations, but the phases are known for tides. A tide model should be included for accurate dynamics. 

\subsubsection{\label{sec:2.2.8}Phobos Tidal Acceleration}

Phobos shows a large acceleration in orbital longitude. Recent fits by Bills et al. \cite{Bills-etal-2005}, Lainey et al. \cite{Lainey-etal-2007}, and Jacobson \cite{Jacobson-2009} give accelerations in the forward orbital longitude $a\,dn/dt$ of $\sim416$ m/yr$^2$. This secular acceleration would easily be detected by PLR giving refined accuracy. The apparent cause of this acceleration is Phobos-raised tides on Mars perturbing Phobos. The tidal bulge raised by Phobos is behind (in time and longitude) Phobos' position sapping energy from the orbit, which consequently shrinks by 3.8 cm/yr. Phobos will eventually impact Mars \cite{Efroimsky-Lainey-2007}. The most important of the tidal components for the secular acceleration should be the second-degree $M_2$ tide of period 5.55 hr on Mars. The small eccentricity (0.015) and inclination (1.1$^\circ$) tend to reduce the influence of other degree 2 tides by $\sim3$ orders of magnitude or more. The influence of tides of higher degree fall off as even powers of $R/a = 1/2.76$, about an order of magnitude per degree. So the degree 3 tide of 3.7 hr period on Mars is a small contributor to the tidal secular acceleration.  

What do tides on Phobos do? Based on evolutionary arguments, Yoder \cite{Yoder-1982} has placed an upper limit on Phobos' $k_2/Q$ of $2\times10^{-7}$, which would make dissipation in Phobos a minor contributor, of order $10^{-3}$ relative to the overall tidal acceleration of Phobos. Periodic tidal displacements on Phobos might reach 1 mm.  

\subsubsection{\label{sec:2.2.9}Impacts}

Mars and other solar system bodies are subject to bombardment by meteoroids. Small new craters are observed on Mars \cite{Malin-etal-2006}. The probability of impact on a body depends on the body's projected area while the recoil from the impulse is inversely proportional to the body's mass. The area to mass ratio for Phobos is $\sim$600 times larger than that for Mars, so impacts are considered here. A 1 kg projectile impacting Phobos at 10 km/s can cause a mean motion change corresponding to $a\Delta n$ up to 0.1 mm/yr along the orbit. Inner solar system impact rates would make a 1 kg or larger impact on Phobos a rare event occurring less than once per century \cite{Neukum-etal-2001,Ivanov-2001,Hartmann-Neukum-2001,Dikarev-etal-2005}. Impacts are not a concern for the dynamics of the PLR mission.  

\subsubsection{\label{sec:2.2.10}Asteroid Masses}

Mars and Earth are perturbed by minor planets in the asteroid belt. Mars is adjacent to the belt and is much more strongly influenced than the Earth. Both analytical and numerical techniques have shown the importance of asteroid perturbations \cite{Williams-1984,Fienga-Simon-2005,Konopliv-etal-2006,Fienga-etal:2009}. The three largest asteroids, 1 Ceres, 2 Pallas and 4 Vesta, are each capable of perturbing Mars by more than one kilometer on time scales of several decades through long-period near resonances. Shorter period perturbations with smaller amplitudes are also effective. For periods of a decade or less, amplitudes are $>100$ m from Ceres and Vesta, somewhat smaller from Pallas, and at least 10 m from four other bodies. There are estimated to be $\sim67$ bodies which can perturb Mars with amplitudes $>1$ m for periods of 10 yr or less and the number is hundreds for several decades. The total perturbation is larger than any single amplitude. Knopoliv et al. found 67 bodies which perturb the Earth-Mars range by $>100$ m long term or $>4$ m short term. 

At a millimeter level Mars is affected by a very large number of small bodies, too many to reliably separate all of the masses during a solution. The simulations of this paper solved for 67 asteroid masses. The asteroid masses affect the simulated solutions for several parameters of interest including the equivalence principle. The acceleration in orbital longitude from solar $GM$ change is difficult to distinguish from periodicities much longer than the data span. Existing Earth-Mars radio range data can be fit well by solving for 20 or more asteroid masses \cite{Konopliv-etal-2006,Folkner-etal-2008,Fienga-etal:2009,Konopliv-etal-2009}, but the asteroids cause the orbit error to grow rapidly for extrapolations into the future. 

Apart from improving the dynamics of the solar system, asteroid masses are of scientific interest. Asteroids have different compositions with physical properties grouped into multiple taxonomic classes indicating a diversity of origin and evolution histories. Their densities are part of the physical properties picture. In addition to fits of radio ranges, existing determinations of asteroid masses come from asteroid-asteroid perturbations observed by optical astrometry \cite{Baer-Chesley-2008}, observations of binary asteroids, several spacecraft flybys and one orbiter. Masses derived from very accurate laser ranges to Mars or Phobos would be sensitive to the more massive asteroids and those that can pass through the inner asteroid belt. The future Gaia astrometric mission is expected to determine masses from perturbations of one asteroid passing another \cite{Mouret-etal-2008}. 

\section{\label{sec:4}Technical Overview of PLR Mission, Experiment, and Methodology}

\subsection{\label{sec:4.1}Mission Description}

The Phobos Laser Ranging mission is based on deploying a laser ranging transponder instrument capable of supporting measurements of distances to Earth with 1 mm accuracy given daily hour-long tracking passes \cite{Murphy-etal-2009}. In our studies, laser ranging in the optical regime was selected rather than radio ranging.  While theoretically it may be possible to achieve mm class accuracy using a dual Ka and X band arrangement, laser ranging is much closer to the state of the art.  Furthermore, ground-based radio antenna calibration would be a serious concern--currently the DSN does not calibrate to even meter accuracy in range. A pulsed, photon counting laser system was selected rather than a coherent system.  The number of received signal photons necessary for coherent detection was not considered attainable at standard laser power levels at interplanetary ranges.

\begin{figure*}[h!]
  \vspace{-7pt}
  \begin{center}
    \includegraphics[width=0.58\textwidth]{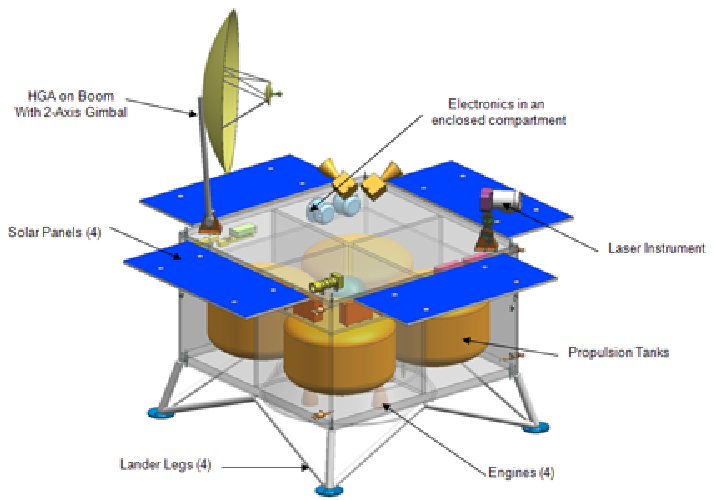}
   \includegraphics[width=0.41\textwidth]{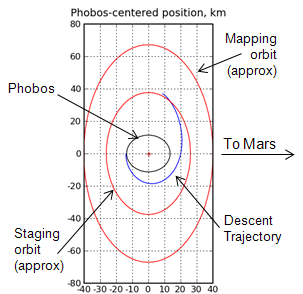}
  \end{center}
  \vspace{-5pt}
  \caption{(a) PLR spacecraft in deployed configuration. (b) Phobos landing trajectory concept.}
\label{fig:sc-concept}
\end{figure*}

The PLR spacecraft configuration is fairly simple. The box-like main structure would have four legs rigidly attached for landing on Phobos (Fig.~\ref{fig:sc-concept}). Given the anticipated characteristics of the spacecraft, it will easily fit within the fairing of an Atlas V-401 launch vehicle, the smallest Mars-capable vehicle under NASA contract for 2015 and beyond. After launch the four small solar panels, high-gain antenna, and laser ranging instrument will be deployed.

Following a 7-month cruise, when the spacecraft arrives at Mars, a propulsive maneuver will place it into a highly elliptical orbit about Mars. This will be followed by a maneuver at apoapsis to raise the periapsis and match the plane of Phobos' orbit, and then a maneuver at peripasis to match the Phobos orbit period (Fig.~\ref{fig:sc-concept}b). Phobos will be observed from orbit for one month with a simple camera to improve the Phobos ephemeris and identify a specific landing location and map gravity field. (The PLR ranging system could be designed to operate in altimeter mode before landing, which may help to improve the determination of the Phobos' gravity field.)
Landing on Phobos will be done semi-autonomously using a laser altimeter (could be separate from PLR transponder) and a feature detection camera similar to those used on the NEAR spacecraft.\footnote{See more details on NEAR spacecrfat at \tt http://www.msss.com/small\_bodies/near\_new/index.html\#top} 

\begin{figure*}[h!]
    \includegraphics[width=0.30\textwidth]{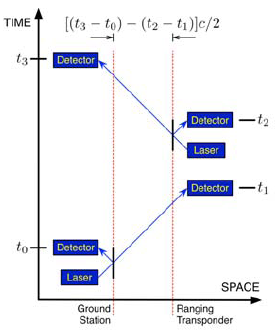}
~~~~~~~
    \includegraphics[width=0.62\textwidth]{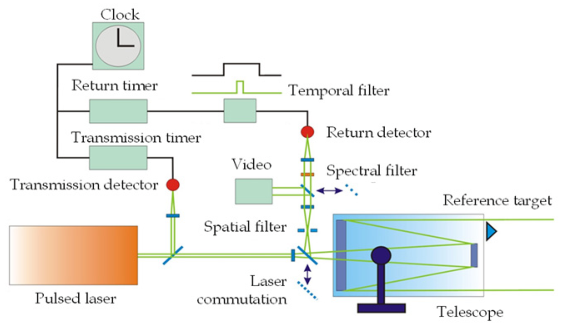}
 \vspace{-0pt}
  \caption{(a) Sequence of laser pulse time-tags used to derive Earth-Phobos round-trip light-time. (b) A block diagram of the laser ranging instrument and Earth instrument for timing transmitted and received pulses on a common detector (see more details on PLR instrument in Fig.~\ref{fig:instrum-block-diag}).} 
\label{fig:observ-concept}
\end{figure*}

After landing, the spacecraft will have approximately three sun and Earth view periods of $>3$ hours, giving more than 5 hours with elevations $>45^\circ$ every Earth day. One of these view periods each day will be used for the hour-long laser ranging experiments. Once each week, a second sunlit period will be used to send the instrument data to Earth via the high-gain antenna. Nominal science operations would continue for 3 years, covering two solar conjunctions. Since there are no expendables (e.g. propellant) during science operations, further observations could occur in an extended mission phase.

\subsection{\label{sec:4.2}Measurement  Description}

The gravity parameter estimation is based on monitoring the light-time between Earth and Phobos. Unlike lunar laser ranging which measures the round-trip light-time from Earth to the moon and back, PLR will use a combination of one-way light-times with asynchronous operation of the laser \cite{Degnan-2002,Birnbaum-Chen-Hemmati-2010}. The advantages of the asynchronous scheme include (for details on laser transponder operation see \cite{Degnan-2002}): (1) noise immunity: the remote station does not transmit pulses in response to numerous background photon detections; (2) stable laser performance: steady pulse generation results in thermally stable conditions; (3) the measurement rates for an ``asynchronous transponder'' are substantially higher compared to an ``echo'' one,  if the probability of detection at one or both ends of the link are small; and (4) multiple Earth stations \cite{Pearlman-etal-ILRS-2002} can range to the remote unit simultaneously without placing conflicting demands on the remote transmitter. The Earth stations will measure the times at which laser pulses are directed toward Phobos and the times of received photons. Similarly, the laser ranging instrument on Phobos will measure the times of pulses transmitted towards Earth and the times of received photons from Earth. Scattered light will be rejected to first order by using a known laser pulse rate and filtering the measured times to find those separated by the signal rate. Timing will be done to an accuracy of 50 ps per photon (though not explicitly synchronized), averaging down to $<$ 3 ps for a typical observation. For the Earth station, the timing accuracy must be kept over the round-trip light time of up to $\sim$3000~s. The Phobos instrument accuracy need only be good over the few minutes it takes to collect enough photons to produce one-millimeter statistics. The timing scheme is indicated in Fig.~\ref{fig:observ-concept}a.

Both the Earth observatory and laser ranging instrument include a telescope and optics to have incoming light and a small fraction of outgoing light illuminate a photon sensitive detector array that records single photon arrivals (Fig.~\ref{fig:observ-concept}b).  The power received by the telescope depends directly on the telescope's collecting area and inversely on the returning spot area.  At the detector, both a pinhole restricting the field of view and a narrow-band-pass filter will be included to reduce background light.

If the combined timing uncertainty per photon is 50 ps (from laser pulse width, detector response, timing electronics, clock), then one needs $(50/3.3)^2 = 225$ photon events in order to establish the one-way distance to 1 mm (3.3 ps) precision. At the estimated rates in the most demanding scenario, this is achieved in a few minutes of ranging. However, it is assumed that all other parameters (beam pointing, divergence) are satisfied, which will not be true all of the time, so hour-long tracking passes are used to ensure that enough valid measurements are obtained.

\begin{table}[h!]
\begin{center}
\vskip -0pt
\caption{Signal and noise estimated for laser ranging from Earth to Phobos and back--most demanding scenario. Note: ``tx'' and ``rx'' stand for transmit and received signal modes. Also, the following notations are used pW$=10^{-12}$W, aW$=10^{-18}$W, pJ$=10^{-12}$J, aJ$=10^{-18}$J. \label{tab:2.1}}
\vskip 1pt
\begin{tabular}{|r|r|r|} \hline  
Link parameters &  Earth to Phobos & Phobos to Earth\\  \hline  
\multicolumn{3}{l}{~~~~~Input parameters:}\\ \cline{1-3} 
wavelength   & 532 nm & 1064 nm \\
tx power & 3 W & 0.25 W \\
tx efficiency & 0.5&0.5\\
tx beam divergence & 25 $\mu$rad & 160 $\mu$rad\\
tx pointing loss & $-$2 dB & $-$2 dB\\
tx atmospheric loss & $-$3 dB & 0 dB\\
tx pulse frequency & 1 kHz & 1 kHz \\
rx atmospheric loss & 0 dB & $-$4.3 dB\\
rx diameter & 0.1 m & 1 m\\
rx efficiency & 0.3&0.3\\
rx field of view & 240 $\mu$rad & 20 $\mu$rad \\
rx detector efficiency & 0.4 & 0.4 \\
background & 65 W/m$^2$/sr/$\mu$m & 20 W/m$^2$/sr/$\mu$m \\
scattered light irradiance & 130 W/m$^2$/sr/$\mu$m & 130 W/m$^2$/sr/$\mu$m \\
Earth's sky irradiance & 0 W/m$^2$/sr/$\mu$m & 100 W/m$^2$/sr/$\mu$m \\
bandpass FWHM & 0.2 nm & 0.2 nm\\
range & 2.6 AU & 2.6 AU\\\hline 
\multicolumn{3}{l}{~~~~~Derived parameters:}\\ \cline{1-3}
photon energy & 0.37 aJ & 0.18 aJ\\
space loss & $-$379.3 dB & $-$373.2 dB\\
rx signal power & 36.1 aW & 37.8 aW\\
planet background power & 0.16 pW & 1.2 pW\\
scattered light power & 17 pW & 2.5 W\\
sky radiance power & 0 pW & 6.3 pW\\
\hline 
\multicolumn{3}{l}{~~~~~Summary results:}\\ \cline{1-3}
incident signal power & 10.8 aW & 1.13 aW\\
incident noise power & 5.1 pW & 3.0 pW\\
SNR & $-$56.8 dB & $-$64.2 dB\\
detected signal rate & 11.6 Hz & 2.4 Hz \\
detected  noise rate & 5.5 MHz & 6.5 MHz\\
10 ns window data volume & 190 Mb/day & 222 Mb/day\\\hline
\end{tabular} 
\end{center}
\vskip -5pt
\end{table}

The laser ranging instrument on Phobos will transmit pulses at a wavelength of 1064-nm for greater power efficiency and lower atmospheric scattered light at the Earth receiver during near sun pointing operations.  A laser pulse rate of 1 kHz will be used with 12 ps pulse-widths and power of 0.25 W. We selected a divergence of 160 $\mu$rad to cover the entire Earth at distances greater than 1 AU. At closer ranges, the instrument must point to a specific area on the Earth. For a 1 m aperture on the Earth station at maximum Earth-Mars range, the average detection rate will be 1.2 photons per second.

Transmission from the Earth station will be at a wavelength of 532 nm and detected by the instrument on Phobos with an efficient, low noise silicon-based single photon detector. The beam transmitted from Earth will have an uplink divergence of 25 $\mu$rad to cover potential pointing and atmospheric seeing variations. Assuming a laser power of 3 W (3 mJ/pulse at 1 kHz repetition rate) and a 12 cm aperture on Phobos, the uplink will deliver 3 detected-photon/s at the maximum Earth-Mars range.

Background light on Earth represents a serious challenge for a detector system capable of single-photon signal detection. Every photon is identical as seen by the detector, so that signal photons may only be differentiated from background photons via time-domain characteristics. Each square arcsec of the illuminated Earth delivers 2000 photons per second into a 0.01 m$^2$ aperture employing a 0.2 nm bandpass filter and 10\% overall detection efficiency. Assuming the detector has a spatial resolution of 1 arcsec, this means that background light will typically dominate the signal in photon count. It becomes clear why echo-transponding is less favorable in these conditions. However, the laser pulses---even if only occasionally detected---will form a regular temporal pattern unlike the background photons, and will therefore be separable \cite{Degnan-2002}. 

Refractive delay of the earth atmosphere imposes a $\sim$2 meter propagation delay, which is routinely removed at the few mm level via surface measurements of pressure, temperature, and humidity \cite{Mendes-etal-2002,Mendes-Pavlis-2004}.

The background event rates--due to scattered sunlight, illuminated body in field-of-view, and the radiance of the Earth atmosphere--will be much greater than signal event rates. At Phobos the laser ranging instrument will search for pulses in a 1 kHz repeat pattern following the tracking passes, and select only pulses within a narrow window about the times of the expected pulses from Earth for transmission. At Earth the photon timings will be matched against predicted pulse rate and Phobos-Earth range to select correct pulses. Adjusting the pulse repetition pattern on Earth to be precisely 1 kHz at Phobos will facilitate this process. Table~\ref{tab:2.1} summarizes a conservative estimate of signal and noise levels at the most distant separation---at conjunction when the sky is also at its brightest (note, ``tx'' and ``rx'' in the Table~\ref{tab:2.1} stand for transmit and received).

\subsection{\label{sec:4.3}System Description}

\subsubsection{\label{sec:4.3.3}Ground Segment System}

One or more Earth observatories will be equipped similar to current lunar laser ranging stations \cite{Pearlman-etal-ILRS-2002} and be outfitted with membrane-type sunlight rejection filters for accurate measurements when Mars is near solar conjunction. The ground stations will use lasers very similar to those on Phobos, with the exception that the frequency will be doubled and the pulse repetition rate will be continuously tunable so that the pulse arrival rate at Phobos is exactly 1 kHz. The detector used in the ground stations will be an InGaAsP hybrid photodiode with a thinned photocathode to reduce timing jitter to less than 40 ps. A solar rejection filter with passband at 532 and 1064 nm must cover the entire $\sim$1 m aperture of the telescope, which is a demanding task, but one that has been previously demonstrated.

\subsubsection{\label{sec:4.3.1}Payload}

\begin{wrapfigure}{R}{0.40\textwidth}
  \vspace{-30pt}
    \includegraphics[width=0.40\textwidth]{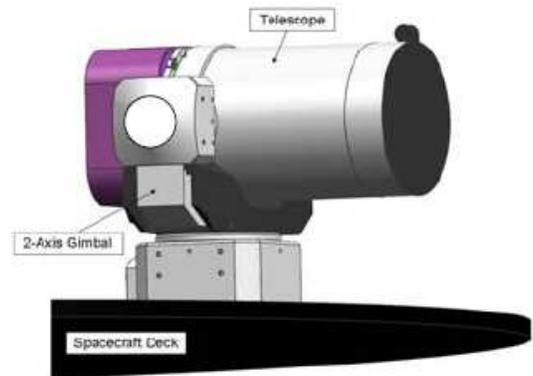}
  \caption{CAD model of laser ranging instrument telescope.}
\label{fig:plrt+gimbal}
  \vspace{20pt}
\end{wrapfigure}

\begin{figure}[h!]
\begin{center}
  \vspace{-5pt}
    \includegraphics[width=0.73\textwidth]{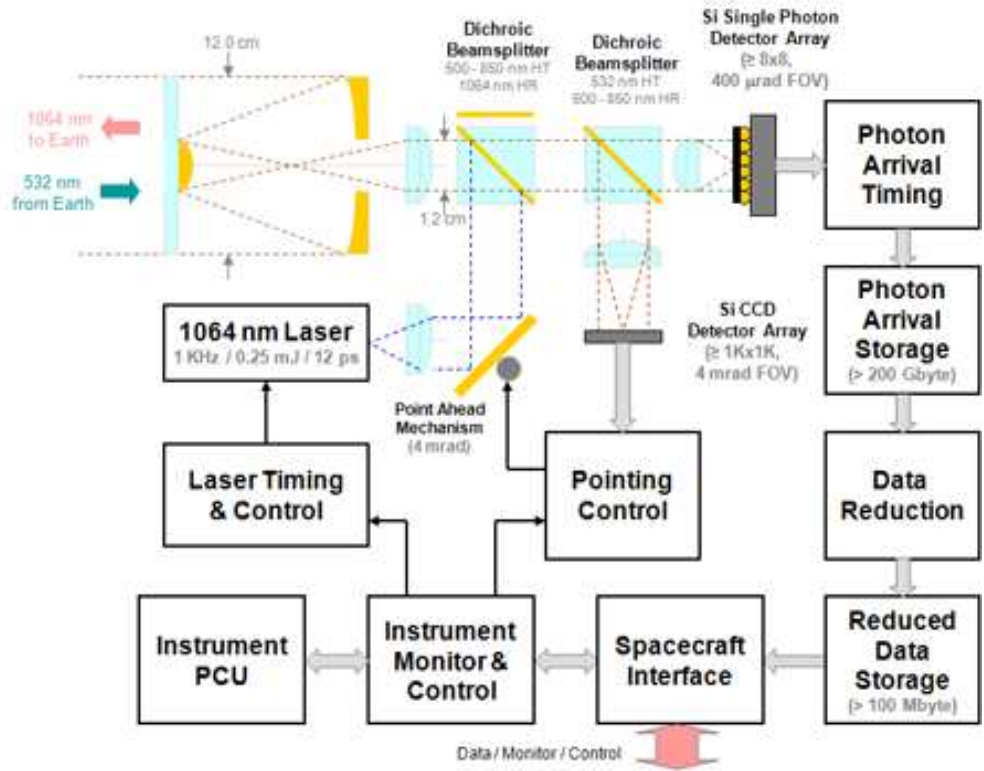}
  \vspace{-5pt}
  \caption{PLR instrument block diagram.}
\label{fig:instrum-block-diag}
  \vspace{-10pt}
\end{center}
\end{figure}

The instrument design is evolved from optical communications terminals under development for interplanetary missions. Prototypes of the communications terminals have been tested in airplanes. For the Phobos laser ranging science objectives, lower pulse rates are acceptable but higher timing resolution is required. The PLR instrument design is based on existing laboratory lasers and array photo-detectors which will be ruggedized to operate in space \cite{Hemmati-etal-2009}. The instrument is shown in Fig.~\ref{fig:plrt+gimbal} and a functional block diagram shown in Fig.~\ref{fig:instrum-block-diag}. The instrument component mass and power are listed in Table~\ref{tab:2.2}.

\begin{table}[h!]
\begin{center}
\vskip -10pt
\caption{Laser ranging instrument mass/power. \label{tab:2.2}}
\vskip 5pt
\noindent
\begin{tabular}{|r|r|r|} \hline  
Component & Mass, kg & Power, W\\  \hline  
\multicolumn{3}{l}{Subassembly: Optics}\\ \cline{1-3}
Primary mirror &0.665 & \\
Secondary mirror &0.018& \\
Sun filter &0.475& \\
Mirror doublets &0.009& \\
Fold mirrors &0.008& \\
Beam splitters &0.062& \\\hline
\multicolumn{3}{l}{Subassembly: Electronics}\\ \cline{1-3}
Power conditioning &3.000 &5.25\\
Controller (FPGA) &3.000 &7\\
Seed laser &0.500 &0.25\\
Pump laser &6.000 &10.5\\
Timing storage &0.500 &2.2\\
GM-APD readout &1.000 &3\\
CCD readout &1.000 &3\\
Gimbal drive &1.000 &10\\
Point-ahead drive &0.250 &1\\
Shutter drive &0.500 &3\\
Clock (USO) &1.000 &2\\
Thermal control &0.500 &2\\\hline
\multicolumn{3}{l}{Subassembly: Mechanical}\\ \cline{1-3}
Telescope gimbal &4.500& \\
Point-ahead actuator &0.065& \\
Shutter activator &0.200& \\
Primary structure &4.800& \\
Secondary structure &0.580& \\\hline
\multicolumn{3}{l}{Subassembly: Thermal}\\ \cline{1-3}
Heat pipe &4.000 &1\\
Insulation &0.500&\\
Survival heaters &0.800&\\
Thermal switches &0.400&\\
Cooler for CCD &0.300 &1 \\
Thermometers &1.000&\\
Thermal strap &2.400&\\
Fine control heaters &0.040 &1\\
Fine control thermometers &0.200&\\\hline
\multicolumn{1}{r}{Total:}&
\multicolumn{1}{r}{39.272}&
\multicolumn{1}{r}{52.200}\\  
\end{tabular} 
\end{center}
\vskip -0pt
\end{table}

\begin{table}[h!]
\begin{center}
\vskip -0pt
\caption{Flight System Mass and Power Estimates. \label{tab:2.3}}
\vskip 0pt
\begin{tabular}{|r|r|r|l|} \hline  
Subsystem &  Mass, kg & Peak Power, W & Peak Power Mode \\ 
  \hline \hline 
Laser Ranging Instrument & 39.3 & 50 & Science\\
Attitude Control & 29.4 & 48 & Landing\\
Command and Data & 22.0 & 47 & Science\\
Power & 71.4 & 71 & Orbit Insertion\\
Propulsion & 170.6 & 217 & Orbit Insertion\\
Structure/Mechanisms & 203.6 & 0& \\
Launch Adaptor & 14.8 & 0& \\
Cabling & 40.4 & 0& \\
Telecom & 21.0 & 64 & Data Transmit\\
Thermal & 38.4 & 165 & Cruise\\\hline 
Spacecraft Total (dry) & 650.9 & 619 & Orbit Insertion\\
Contingency & 280.1 & 266 & \\
Spacecraft + Contingency & 931.0 & 885 & \\
Propellant & 1216.2 &  & \\\hline 
\multicolumn{1}{r}{Spacecraft Total (wet):}&
\multicolumn{1}{r}{2147.2} &  
\multicolumn{2}{l}{}\\   
\end{tabular} 
\end{center}
\vskip -15pt
\end{table}

The instrument consists of a gimbaled 12 cm telescope, a 250 mW Nd:YAG laser operating at 1064 nm, a silicon Geiger-mode (GM) avalanche photodiode detector (APD) array in $8\times$8 format, a multiplexed ASIC timing system to go with the detector, a clock, and an acquisition/pointing CCD camera with thermo-electric cooler. A small actuated mirror will impart a point-ahead angle to the beam with respect to the receiver direction of up to 330 $\mu$rad, which will be monitored via the centroid of a sample of the transmit beam imaged onto the CCD. A small fraction of the outgoing laser light will be reflected back from a reference mirror to the single-photon detector array for timing transmitted pulses. The laser will derive from a Semiconductor Saturable Absorber Mirror (SESAM) fiber seed laser, delivering 12 ps pulses at a 50 MHz repetition rate, followed by a pulse-picker that will accept only every 50,000th pulse to form a 1 kHz train. A laser diode-pumped regenerative amplifier will then deposit $\sim$250 $\mu$J of energy into each pulse for transmission to Earth.

The instrument electronics includes timing, pointing control, data storage, and a data post-processor. After acquiring photon times for the one hour tracking pass, the post-processor will look for the 1 kHz transmit pulse pattern from Earth, delete times outside a 10 ns window about that estimated rate, and format times into differences for reducing data volume. Edited times will be collected for transmission to Earth via radio, with a data volume of 5 Mb per day.

\subsubsection{\label{sec:4.3.2}Flight System}

The PLR spacecraft includes all equipment necessary to deliver and support the laser ranging instrument on Phobos. The spacecraft structure is a simple box-like construction to accommodate the payload, spacecraft electronics, propulsion, attitude control, telecommunications, thermal, and power subsystems. The mass by subsystem is given by Table~\ref{tab:2.3}. There are several operating modes; a typical power by subsystem is also given in Table~\ref{tab:2.3}. With the Atlas V 401 as the launch vehicle, mass and volume capabilities are much greater than necessary, so subsystems have been optimized to minimize cost rather than mass. Selected redundancy has been included in subsystem design to reduce risk where cost effective.

Attitude control will be done using thrusters since there is no stringent pointing requirement for the spacecraft. Attitude control sensors include standard star trackers, inertial measurement units, coarse sun sensors, optical navigation camera, terminal descent camera, terminal descent laser altimeter (i.e., a NEAR-type altimeter mentioned earlier), and electronics for driving the laser range instrument pointing gimbal. 

The command and data handling subsystem is based on the architecture being developed for the Mars Science Laboratory (MSL). It includes RAD750 processors, interfaces to the attitude control, telecommunication, power, and payload subsystems, with sufficient memory storage to accumulate up to two weeks of instrument data.

The power subsystem includes 5.3 m$^2$ solar panels and batteries to operate in absence of sun-light. The battery size is dictated by the propulsion system operation during Mars orbit insertion.

Propulsion will be done using a mono-propellant to minimize cost of tanks and thrusters. Standard thrusters of 300 N(4), 90 N(4), 4.5 N(4), and 0.9 N(12) will be used for Mars orbiter insertion, attitude control, and landing on Phobos. Four propellant tanks will contain adequate propellant with margin. After landing the remaining propellant will be vented.

Thermal control will be done using standard multi-layer insulation and heaters. Spacecraft and electronics will be located in a thermally isolated box that will be kept warm during dark periods on Phobos. The propellant tanks will be kept warm until landing and uncontrolled afterwards.

Telecommunications will utilize standard X-band transponders and traveling-wave tube amplifiers, with a deployed pointable 1.5-m diameter high-gain antenna for sending instrument data to Earth, and two low-gain antennas used for contacts during cruise.

\subsubsection{\label{sec:4.3.4}Mission Operations System}

Science operations will be done for one-hour tracking passes each Earth day. The spacecraft will have three view periods each day to select for operations. The tracking pass time will be selected to accommodate a view of the selected ground observatory for that day. Observing times will be optimized to have the observatory view Phobos at a high elevation, and sample the Phobos orbit by varying the time to alternate between before and after maximum elevation at the spacecraft. Times will be selected one week in advance. Once per week the spacecraft will communicate with the Deep Space Network to download the previous week of science data and to upload the schedule of tracking times for the next week.

\subsection{\label{sec:4.4}Science Data Collection, Analysis, and Archive}

The primary data product from PLR will be a series of timestamps for photon events, each also indicating pixel of origin on the detector. The signal-finding algorithms will permit sorting the local fiducial return (launch times) from one or more receive pulse trains from stations on the earth. Compressed statistics on background rate for each pixel will also be available. To the extent permitted by communications capabilities, full-rate data (including background events) will sometimes be included so that it will be possible to verify on-board compression efficacy. Ground stations will likewise present timestamps for their photon launch and receive events. Auxiliary data such as meteorological conditions at each site, system health parameters, pointing information, etc. will join the total dataset, but will constitute a small fraction of the data volume.

Data processing occurs in two stages. First, the event times are combined into meaningful ``normal points,'' comprised of a representative midpoint time and round-trip time constructed from the two one-way measurements. During this process, the Phobos clock is calibrated (frequency offset measured), and various detector and timing system performances are verified. Second, having a set of normal points from a variety of earth stations, the measurements are compared to a sophisticated, parameterized physical model of the solar system, including gravitational dynamics and planetary physics influences. The difference between the model and actual ranges is minimized with respect to the parameters in the model--many of which are simply initial conditions for solar system bodies. The model for gravity allows exploration of departures from general relativity, which is the primary focus of the PLR project.

\section{\label{sec:concl}Conclusions}

Interplanetary laser ranging may offer very significant improvements in many areas including fundamental physics, planetary research and deep-space navigation.  What is critical for the purposes of fundamental physics is that, while in free space, ILR allows for a very precise trajectory estimation to an accuracy of less than 1 cm at distances of $\sim 2$ AU. With the recent successful laser transponder experiments conducted with MLA and MOLA instruments \cite{Smith-etal-2006,Sun-etal-2006,Abshire-etal-2006,Degnan-2007a}, ILR rapidly becoming a mature technology.  A mm-level ranging precision over interplanetary distances is within reach, thus opening a way for significant advances in the tests of gravity on solar system scales \cite{Degnan-2007b}.

Laser ranging to Phobos with mm-level range sensitivity would allow major advances in the tests of fundamental gravity \cite{Turyshev-2008}.  A relativistic time delay of the laser signal passing through the Sun's gravity field will deliver a high accuracy determination of the Eddington parameter $\gamma$.  It is important to improve the accuracy of tests of the parameter $\gamma$ since any deviation from general relativity would signal new physics.  Although, no break-down of the PPN formalism is expected, the unprecedented precision of PLR could point out where the effects $\propto G^2$ on light propagation kick in. We emphasize that PLR offers the opportunity to significantly improve the measurement of the time-rate-of-change in the gravitational constant $G$, and to search for a new long-range interaction via tests of the Yukawa forces on interplanetary scales.  In addition, PLR will test the EP looking for Jupiter-induced deviations in the Earth-Phobos range as these bodies orbit the Sun.

In addition to the primary astrophysics science, a variety of other scientific advances would be made by the PLR mission.  For Phobos, these include detailed gravity, topographical, and thermographic maps, unprecedented orbital determination accuracy and knowledge of physical libration characteristics, an upper limit on tidal behavior, and local information about the landing site, such as rock distribution, regolith characterization, etc. For Mars, its overall body distortion and dissipation due to tides could be measured. 

The future deployment of laser transponders on interplanetary missions will provide new opportunities for highly improved tests of gravitational physics.  With their anticipated capabilities, interplanetary transponders will lead to very robust advances in the tests of fundamental physics and could discover a violation or extension of general relativity, or reveal the presence of an additional long range interaction in physical laws. As such, these devices should be used in planning both the next steps in lunar exploration and future interplanetary missions to explore the solar system.

\begin{acknowledgements}
The work described in this report was performed at the Jet Propulsion Laboratory, California Institute of Technology, under a contract with the National Aeronautics and Space Administration. Government sponsorship acknowledged. The anonymous referee is thanked for valuable
suggestions that improved the text and made it more
readable.
\end{acknowledgements}

\bibliography{plr-paper}

\end{document}